\def\hpec{true}

\ifdefined\hpec         

  \documentclass[conference]{IEEEtran}

\else                   

  \documentclass[12pt,draft]{article}

  \def\usenumberlines{true}

\fi

\makeatletter
\def\endthebibliography{%
  \def\@noitemerr{\@latex@warning{Empty `thebibliography' environment}}%
  \endlist
}
\makeatother

\def\BibTeX{{\rm B\kern-.05em{\sc i\kern-.025em b}\kern-.08em
    T\kern-.1667em\lower.7ex\hbox{E}\kern-.125emX}}

\ifdefined\hpec         
  \usepackage{balance}
\else                   
  \usepackage[margin=1in]{geometry}
  \usepackage{setspace}
\fi

\usepackage{amsmath}
\usepackage{amsfonts}
\usepackage{amssymb}
\usepackage{amsthm}
\usepackage{mathrsfs}
\usepackage{array}

\usepackage{float}
\usepackage[font=small,labelfont=bf,hypcap=true]{caption}

\usepackage{graphicx,subfigure}
\usepackage[hyperref,x11names,cmyk,pdftex]{xcolor}
\usepackage{tikz}
\usepackage{float}
\usepackage{subfloat}
\usepackage{tablefootnote}
\usepackage{adjustbox}

\usepackage{multirow}
\usepackage{booktabs}

\usepackage{enumerate}
\usepackage{paralist}

\usepackage[bottom]{footmisc}

\usepackage[super]{nth}

\usepackage{helvet}
\usepackage{textcomp}
\usepackage[utf8]{inputenc}
\usepackage[resetfonts]{cmap}

\usepackage[layout={inline},author=]{fixme}

\usepackage{algorithm,algorithmic}

\usepackage{csquotes}
\ifdefined\hpec
	\usepackage[pdftex,bookmarks=false]{hyperref}
	\usepackage{cite}
\else                   
  \usepackage[pdftex,breaklinks=true,bookmarks=true]{hyperref}

  \ifdefined\usenumberlines

    \usepackage{lineno}
    \linenumbers\relax

    \newcommand*\patchAmsMathEnvironmentForLineno[1]{%
      \expandafter\let\csname old#1\expandafter\endcsname\csname #1\endcsname
      \expandafter\let\csname oldend#1\expandafter\endcsname\csname end#1\endcsname
      \renewenvironment{#1}%
      {\linenomath\csname old#1\endcsname}%
      {\csname oldend#1\endcsname\endlinenomath}}%
    \newcommand*\patchBothAmsMathEnvironmentsForLineno[1]{%
      \patchAmsMathEnvironmentForLineno{#1}%
      \patchAmsMathEnvironmentForLineno{#1*}}%
    \AtBeginDocument{%
      \patchBothAmsMathEnvironmentsForLineno{equation}%
      \patchBothAmsMathEnvironmentsForLineno{align}%
      \patchBothAmsMathEnvironmentsForLineno{flalign}%
      \patchBothAmsMathEnvironmentsForLineno{alignat}%
      \patchBothAmsMathEnvironmentsForLineno{gather}%
      \patchBothAmsMathEnvironmentsForLineno{multline}%
    }

  \fi

\fi

\title{\LARGE Damping Effect on PageRank Distribution}

\ifdefined\hpec
\author{%
  \IEEEauthorblockN{%
    Tiancheng Liu,
    Yuchen Qian,
    Xi Chen, and
    Xiaobai Sun
    \\ 
    \IEEEauthorblockA{%
    \begin{tabular}{c}
      Department of Computer Science, Duke University, Durham, NC 27708, USA
    \end{tabular}%
  }%
}}

\else
\fi 

\newcommand{\pdfauthors}{%
  T. Liu, Y. Qian, X. Chen, X. Sun}

\hypersetup{%
  pdffitwindow=false,%
  pdfstartview={FitH},%
  colorlinks,%
  pdfauthor={\pdfauthors},%
  pdftitle={Damping Effect on PageRank Distribution},%
  citecolor=PaleGreen4,%
  filecolor=DarkOrchid4,%
  linkcolor=OrangeRed4,%
  urlcolor=black%
}

\setkeys{Gin}{draft=false}

\pdfpageattr{/Group <</S /Transparency /I true /CS /DeviceRGB>>}

\let\leftorig\left
\let\rightorig\right
\renewcommand{\left}{\mathopen{}\mathclose\bgroup\leftorig}
\renewcommand{\right}{\aftergroup\egroup\rightorig}

\fxsetup{inline}
\fxusetheme{color}

\hypersetup{final}

\makeatletter
\DeclareRobustCommand\ttfamily{%
  \not@math@alphabet\ttfamily\mathtt
  \fontfamily\ttdefault\selectfont\hyphenchar\font=-1\relax}
\makeatother
\DeclareTextFontCommand{\tturl}{\ttfamily\hyphenchar\font=`/}

\begin{document}


\ifdefined\hpec\else
  \doublespacing
\fi


%
\ifdefined\hpec
  \maketitle

%
\else


  \pdfbookmark[1]{Title}{sec:title}
  \maketitle


  \phantomsection
  \pdfbookmark[1]{Contents}{sec:contents}
  \tableofcontents
  \clearpage

\fi



\begin{abstract}
  This work extends the personalized PageRank model invented by Brin and
  Page to a family of PageRank models with various damping schemes. The
  goal with increased model variety is to capture or recognize a larger
  number of types of network activities, phenomenons and
  propagation patterns. The response in PageRank distribution
  to variation in damping mechanism is then characterized analytically,
  and further estimated quantitatively on $6$ large real-world link
  graphs. The study leads to new observation and empirical findings. It
  is found that the difference in the pattern of PageRank vector
  responding to parameter variation by each model among the $6$ graphs is
  relatively smaller than the difference among $3$ particular models
  used in the study on each of the graphs. This suggests the utility of
  model variety for differentiating network activities and propagation
  patterns. The quantitative analysis of the damping mechanisms over 
  multiple damping models and parameters is facilitated by a highly efficient algorithm,
  which calculates all PageRank vectors at once via a commonly shared,
  spectrally invariant subspace. The spectral space is
  found to be of low dimension for each of the real-world graphs.
\end{abstract}




\section{Introduction}
\label{sec:intro}


Personalized PageRank, invented by Brin and Page~\cite{page1999pagerank,
  ilprints361}, revolutionized the way we model any particular type of
activities on a large information network. It is also intended to be
used as a mechanism to counteract malicious manipulation of the
network~\cite{page1999pagerank, ilprints361, sheldon2010manipulation}. 
PageRank has underlain Google's
search architecture, algorithms, adaptation strategies and ranked page
listing upon query.  It has influenced the development of other
search engines and recommendation systems, such as topic-sensitive
PageRank\cite{haveliwala2003topic}. Its impact reaches far beyond
digital and social networks. For example, GeneRank is used for
generating prioritized gene
lists\cite{morrison2005generank,wu2010krylov}. The seminal
paper~\cite{page1999pagerank} itself is directly cited more than ten
thousands times as of today.  As surveyed in
\cite{langville2004deeper,berkhin2005survey}, a lot of efforts were made
to accelerate the calculation of personalized PageRank vectors, in part
or in whole \cite{ilprints596,ilprints579}. Certain investigation were
carried out to assess the variation in PageRank vector in response to
varying damping
parameter~\cite{boldi2005pagerank,bressan2010choose}. Most efforts on
PageRank study, however, are ad hoc to the Brin-Page model. Chung made
a departure by introducing a diffusion-based PageRank model and applied
it to graph cuts~\cite{chung2007heat,chung2009local}.

In this paper we follow Brin and Page in the modeling aspect that
warrants more attention as the variety of networks and activities on the
networks increases incessantly.  We extend the model scope to capture
more network activities in a probabilistic sense. We study the damping
effect on PageRank distribution.  We consider the holistic distribution
because it serves as the statistical reference for inferring conditional
page ranking upon query. Our study has three intellectual merits with 
practical impact.
\begin{inparaenum}[(1)]
\item A family of damping models, which includes and connects the
  Brin-Page model and Chung's model. The family admits more
  probabilistic descriptions of network activities. 
\item A unified analysis of damping effect on personalized PageRank
  distribution, with parameter variation in each model and comparison
  across models. The analysis provides a new insight into the solution
  space and solution methods.
\item A highly efficient method for calculating the solutions to all
  models under consideration at once, particular to a network and a
  personalized vector.  Our quantitative analysis of $6$ real-world
  network graphs leads to new findings about the models and networks under
  study, which we present and discuss in Sections~\ref{sec:theory-analysis}
  and \ref{sec:numerical-experiments}.
\end{inparaenum} 
Our modeling and analysis methods can be potentially used for
recognizing and estimating activity or propagation patterns on a
network, provided with monitored data.
%
%

 


\section{PageRank models}
\label{sec:pageRank-models}
%


We first review briefly two precursor models and then introduce a family
of PageRank models.

\subsection{Brin-Page model} 
\label{subsec:Brin-Page-model}
%

Brin and Page describe a network of webpages as a link graph, which is
represented by a stochastic matrix $P$~\cite{page1999pagerank}. We adopt
the convention that $P$ is stochastic columnwise. Every webpage is a
node with (outgoing) links, i.e., edges, to some other webpages and with
incoming edges or backlinks as citations to the page. If page $j$ has
$n_j >0 $ outgoing links, then in column $j$ of $P$, $P_{i,j} = 1/n_j$
if page $j$ has a link to page $i$; $P_{i,j}=0$, otherwise. In row $i$
of $P$, every nonzero element $P_{i,j}$ corresponds to a backlink from
$j$ to $i$.
%
In the Brin-Page model, the web user behavior is described as a random
walk on a personalized Markov chain (i.e., a discrete-time Markov chain)
associated with the following probability transition matrix
\begin{equation}
\label{eqn:bp-Bernoulli}
M_{\alpha}(v) = \alpha P + (1 - \alpha)v e^{\rm T}, 
\quad \alpha \in (0,1), 
\quad e^{\rm T}v = 1, 
\end{equation}%
where $v\geq 0$ is a personalized or customized distribution/vector, $e$
denotes the vector with all elements equal to $1$, and the \emph{damping
  factor} $\alpha$ describes a Bernoulli decision process.  At each
step, with probability $\alpha$, the web user follows an outlink; or
with probability $(1-\alpha)$, the user jumps to any page by the personalized
distribution $v$.  The personalized transportation term is
innovative. It customizes the Markov chain with respect to a particular
type of relevance.
In this paper we assume that the personalized vector $v$ is given and
fixed, and focus on investigating the damping effect. In particular, 
with Brin-Page model, we focus on the role of $\alpha$.
The notation for the Markov chain may thus be simplified to
$M_{\alpha}$, or $M$ when $\alpha$ is clear from the context.

Page ranking upon a search query depends on the stationary PageRank
distribution, denoted by $x=x(\alpha)$, of the Markov chain:
\begin{equation}
\label{eqn:ppr-markov-equi}
  M_{\alpha}  \,  x = x, 
  \quad e^{\rm T}x = 1. 
\end{equation}%
Arasu et al\cite{arasu2002pagerank} recast the eigenvector equation
(\ref{eqn:ppr-markov-equi}) to a linear system to solve for $x$,
\begin{equation}
\label{eqn:ppr-linear}
(I - \alpha P)x = (1 - \alpha)v, 
\end{equation}
where $I$ is the identity matrix.
Because $\alpha \|P\|_{1} < 1$, the solution can be expressed via the 
Neumann series for the inverse of $(I-\alpha P)$, 
\begin{equation}
\label{eqn:bp-solution}
x(\alpha)  = (1-\alpha) \sum_{k = 0} ^ {\infty} \alpha^k P^k v.
\end{equation}
The weights $(1-\alpha)\alpha^k$ decrease by the factor $\alpha$ from one step to the
next. In \cite{page1999pagerank}, Brin and Page set $\alpha$ to $0.85$.

In (\ref{eqn:bp-solution}), the solution to Brin-Page model with
$\alpha \in (0,1)$ is not the stationary distribution of network $P$. It
is analyzed in terms of steps on $P$. Term $k$ represents the probabilistic
accumulation of the Bernoulli decision process at each step by
(\ref{eqn:bp-Bernoulli}) to step $k$, $k\geq 0$.  Every step has its
print in distribution $x(\alpha)$.


%

\subsection{Chung's model} 
\label{subsec:Chung-model}

Chung introduced a PageRank model~\cite{chung2007heat} in the form of a
heat or diffusion equation, with $v$ as the initial distribution,
\begin{equation}
\label{eqn:heat-model}
  \frac{\partial x}{\partial \beta} = - (I - P)x, \quad x(0) = v,
\end{equation}
where $( I - P ) $ is the Laplacian of the link graph, and we use
$\beta$ to denote the time variable.
From the viewpoint of probabilistic theory, model (\ref{eqn:heat-model})
is underlined by the Kolmogorov's backward equation system for a continuous-time
Markov chain with $-(I-P)$ as the transition rate matrix and with the
identity matrix $I$ as the initial transition matrix.
The solution to (\ref{eqn:heat-model}) is 
\begin{equation}
\label{eqn:heat-solution}
x(\beta) 
= e^{-\beta(I-P)}v 
= e^{-\beta} \sum_{k = 0} ^ {\infty} \frac{\beta^k}{k!} P^k v. 
\end{equation}
%



\subsection{A model family}
\label{subsec:model-family}
%

We introduce a family of PageRank models.  
Each member model is characterized by a scalar \emph{damping
  variable} $\rho$ and a {\em discrete} probability mass function (pmf)
$ w(\rho) = \{ w_k=w_k(\rho), \ k \in \mathbb{N}_w \} $. The support
$ \mathbb{N}_{w} \subset \mathbb{N} $ may be finite or infinite.  There
are a few equivalent expressions to describe our models.
We start by defining the model with a kernel function $f(\lambda, \rho)$, 
\begin{equation}
\label{eqn:uni-random-walk} 
f(\lambda, \rho) = \sum_{k \in N_w} w_k(\rho)\, \lambda^k, 
\quad | \lambda | \leq 1.   
\end{equation}%
The solution specific to network graph $P$ and personalized vector $v$ is, 
\begin{equation}
\label{eqn:uni-random-walk} 
x_f(\rho)   = f(P)v = \left(\sum_{k \in N_w} w_k(\rho)\, P^k \right) v . 
\end{equation}%
The matrix function $f(P)$ is stochastic. The rank
distribution vector $x_f$ is the superposition of step terms with
probabilistic weights $w_k$. The step term $k$ describes the
probabilistic propagation of $v$ at step $k$.
For a specific case, the damping variable may have a designated label,
with a specific range, and the pmf may have a specific support and
additional parameters.  For convenience, we assume $\mathbb{N}$ as the
support. Over the infinite support, the damping weights must decay after
certain number of steps and vanish as $k$ goes to infinity. In theory, 
every discrete pmf can be used as a model kernel in (\ref{eqn:uni-random-walk}).
In practice, each describes a particular type of activity or propagation.

The family includes Brin-Page model and Chung's model.  For the former,
the damping variable is denoted by $\alpha$, the damping weights
$(1-\alpha)\alpha^k$, $k\geq 0$, follow the geometric distribution with
the expected value $\alpha(1-\alpha)^{-1}$. The kernel function
is $ (1-\alpha)(1-\alpha\lambda)^{-1}$.  For Chung's model, we
denote the damping variable by $\beta$, $\beta > 0$. The damping weights
$e^{-\beta}\beta^k/k!$, $k\geq 0$, follow the Poisson distribution with
the expected value $\beta$. The model's kernel function is
$e^{-\beta(1-\lambda)}$.

We describe a few other models, among many, in the family. In fact, the
precursor models are two special cases of the model associated with the
Conway-Maxwell-Poisson (CMP) distribution, which has an additional
parameter $\nu$ to the pmf,
\begin{equation} 
\notag
\label{eqn:CMP-weights}
  w_k( \rho, \nu) 
= \frac{\rho^{k}}{(k!)^{\nu}\, Z}, 
 \quad \nu > 0, 
\end{equation}
where $Z$ is the normalization scalar, and $\nu$ is the decay
rate parameter. The case with $\nu=0$ is the geometric distribution; the
case with $\nu=1$ is the Poisson distribution. If the value of $\rho$ is
fixed, the weights decay faster with a larger value of $\nu$.
The negative binomial distribution, or the Pascal distribution, also
includes the geometric distribution as a special case. It includes
other cases that 
render damping weights with slower decay rates.

In the rest of the paper, for the purpose of including and illustrating
new models, we use the model associated with the logarithmic
distribution, for $\gamma \in (0,1)$, 
\begin{equation}
\label{eqn:log-gamma} 
 f(\lambda,\gamma) =
    \frac{-1}{\ln(1 - \gamma)} 
    \sum_{k = 1}^{\infty} \frac{(\gamma\lambda)^{k}}{k}
   = \frac{\ln(1 - \gamma\lambda)}{\ln(1 - \gamma)}. 
\end{equation}%
The weights decrease slightly faster than the geometrically distributed
ones, but not in the CMP distribution class. 

We now present the system of linear equations with $x(\rho)$ in 
(\ref{eqn:uni-random-walk}) as the solution, 
\begin{equation}
\label{eq:model-equation} 
A(P)x = v, 
\quad A(P) = f^{-1}(P). 
\end{equation}%
The matrix $A$ is an $M$ matrix. In particular, 
$A = (1-\alpha)^{-1}(I - \alpha P)$ for Brin-Page model,
$A = e^{\beta (I - P)}$ for Chung's model and
$A= \ln(1-\gamma) \, \ln^{-1}(I - \gamma P) $ for the log-$\gamma$
model (\ref{eqn:log-gamma}).
The algebraic model expression (\ref{eq:model-equation}) will be used
next for the model expression in a differential equation.


%

 
%


\section{Response to variation in damping}
\label{sec:theory-analysis}
%

We provide a unified analysis of the response in PageRank distribution
to the variation in the damping parameter value as well as to the
change, or connection, from one model to another.

\subsection{Intra-model damping variation}
\label{subsec:pagerank-sensitive}

By (\ref{eqn:uni-random-walk}), we obtain the trajectory of the PageRank
vector $x(\rho)$ with the change in the damping variable $\rho$,
\begin{equation}
\label{eqn:trajectory-equation}
 \dot{x}(\rho) = \frac{d x(\rho)}{d \rho} 
 = \frac{\partial }{\partial \rho}f(P) v = Q(P) \, x(\rho), 
\end{equation}%
where $Q(P) = \frac{\partial }{\partial \rho}f(P) f^{-1}(P)$ by
(\ref{eq:model-equation}), which we may refer to as the $\rho$-transition matrix.
Equation (\ref{eqn:trajectory-equation}) generalizes Chung's diffusion
model (\ref{eqn:heat-model}), in which the $\beta$-transition matrix
$Q=-(I-P)$ is independent of $\beta$.
For the Brin-Page model with damping variable $\alpha$,
\begin{equation}
\label{eqn:bp-trajectory-equationpde}
  Q(\alpha) = 
  \left[ P(I - \alpha P)^{-1} - (1-\alpha)^{-1} I\right]. 
\end{equation} 
For the log-$\gamma$ model (\ref{eqn:log-gamma}),  
\begin{equation}
\label{eqn:bp-trajectory-equationpde}
   Q(\gamma) =  
   \frac{(1-\gamma)^{-1}}{\ln(1 - \gamma)} I 
         - P(I-\gamma P)^{-1} (\ln(I - \gamma P))^{-1}. 
\end{equation} 
For each model, $e^{\rm T}Q=0$. 




In addition to the element-wise response in the rank vector, we would
also like to have an aggregated measure of the response to variation in
$\rho$. Let $x(\rho_{o})$ be a reference PageRank vector.  We
may use Kullback-Leibler divergence \cite{kullback1951information} to measure the discrepancy of
$x(\rho)$ from $x(\rho_{o})$,
\begin{equation}
\label{eqn:kl}
  KL(x(\rho), x(\rho_{o}) ) = 
   \sum_{i} x_i(\rho) \log {\frac{x_i(\rho)}{x_i(\rho_{o})}} . 
\end{equation}
When $\rho=\rho_{o}$, $KL(x(\rho), x(\rho_{o})) = 0$. 
We have the rate of change in KL divergence with the variation in
$\rho$, 
\begin{equation}
\label{eqn:kl-derivative}
  \frac{d}{d \rho} KL(x(\rho), x(\rho_{o})) 
 =  \dot{x}(\rho)^{\rm T} 
    \left( \log {x(\rho)} -\log{x(\rho_{o})}  + e \right)
\end{equation}
We will describe in Section~\ref{sec:batch-ranking} efficient
algorithms for calculating the vectors and measures above.



\subsection{Inter-model correspondence}
\label{subsec:damping-corres}

Each model has its own damping form and parameter. The expected value of
the step weight distribution is, 
\begin{equation}
\label{eqn:expect-steps}
\mu(w(\rho)) = \sum_{k \in \mathbb{N}_{w}} k \cdot w_k(\rho). 
\end{equation}
We may explain this as the expected value of walking steps.  We
establish the point of correspondence between models by their expected
values. That is, for any two models, we set their expected values equal
to each other. Without loss of generality, we let the expected values
for the Brin-Page model serve as the reference. In particular, we have
the correspondence equalities
\begin{equation}
\label{eqn:param-corres}
  \frac{\alpha}{1-\alpha} = \beta, 
  \quad 
  \frac{\alpha}{1-\alpha} 
 = \left(\frac{\gamma}{1 - \gamma}\right)\frac{-1}{\ln(1 - \gamma)}
\end{equation}%
for Chung's model and for the log-$\gamma$ model, respectively. We will show
the comparisons in PageRank vectors at such correspondence points in
Section~\ref{sec:numerical-experiments}.




%


\section{Efficient algorithms for batch ranking}
\label{sec:batch-ranking}
%
%


We introduce novel algorithms for efficient quantitative analysis of
damping effect on PageRank distribution.  Provided with a network graph
$P$ and a personalized distribution vector $v$, the algorithms can be
used in one batch of computation across multiple models as well as over
a range of damping parameter value per model.

\subsection{Reduction to irreducible subnetworks}
\label{subsec:model-reduction}
%

\begin{figure}[!htb]
  \centering
  \subfigure[original link graph]{
    \includegraphics[width=0.22\textwidth]{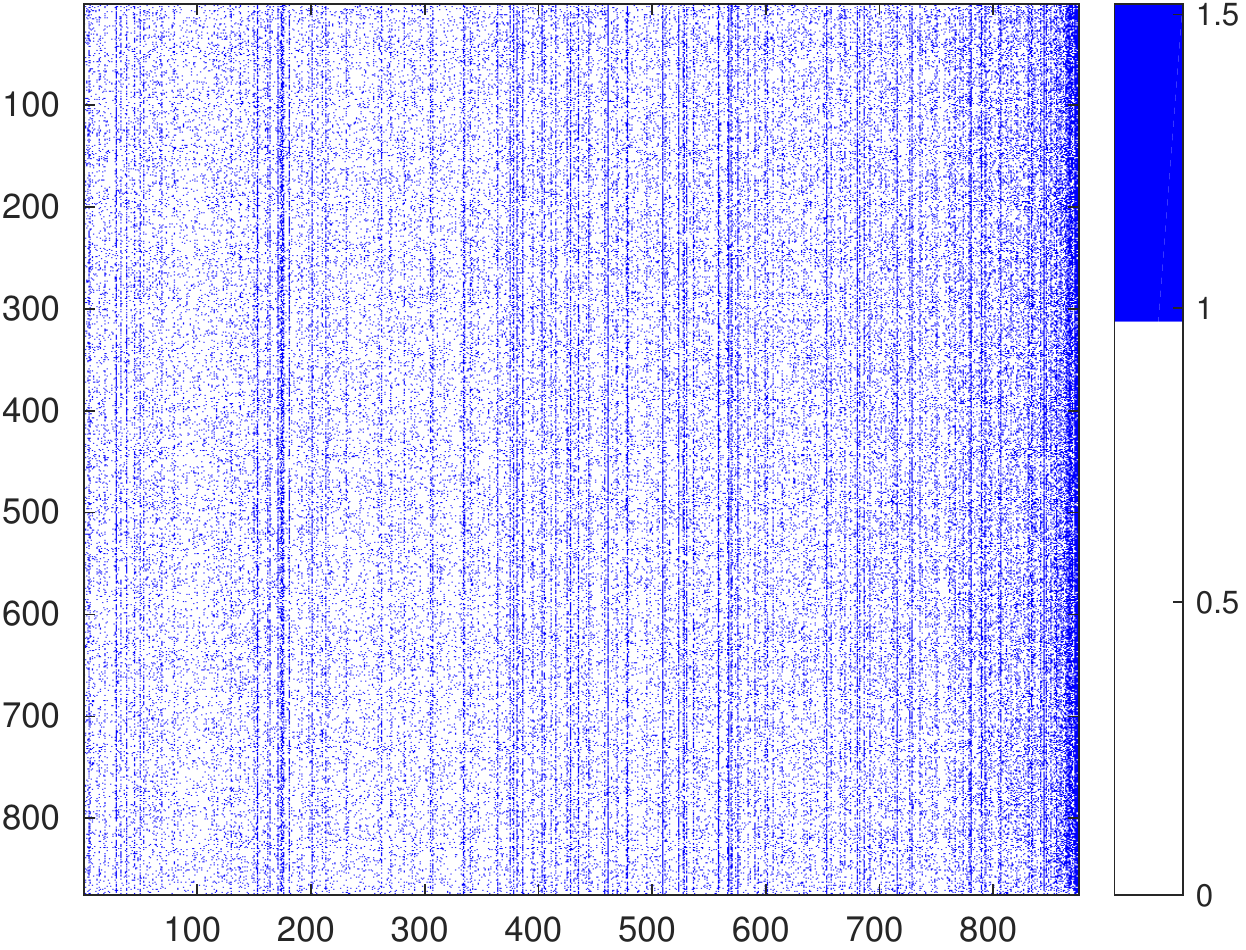}
  }
  \subfigure[link graph after DM permutation]{
    \includegraphics[width=0.22\textwidth]{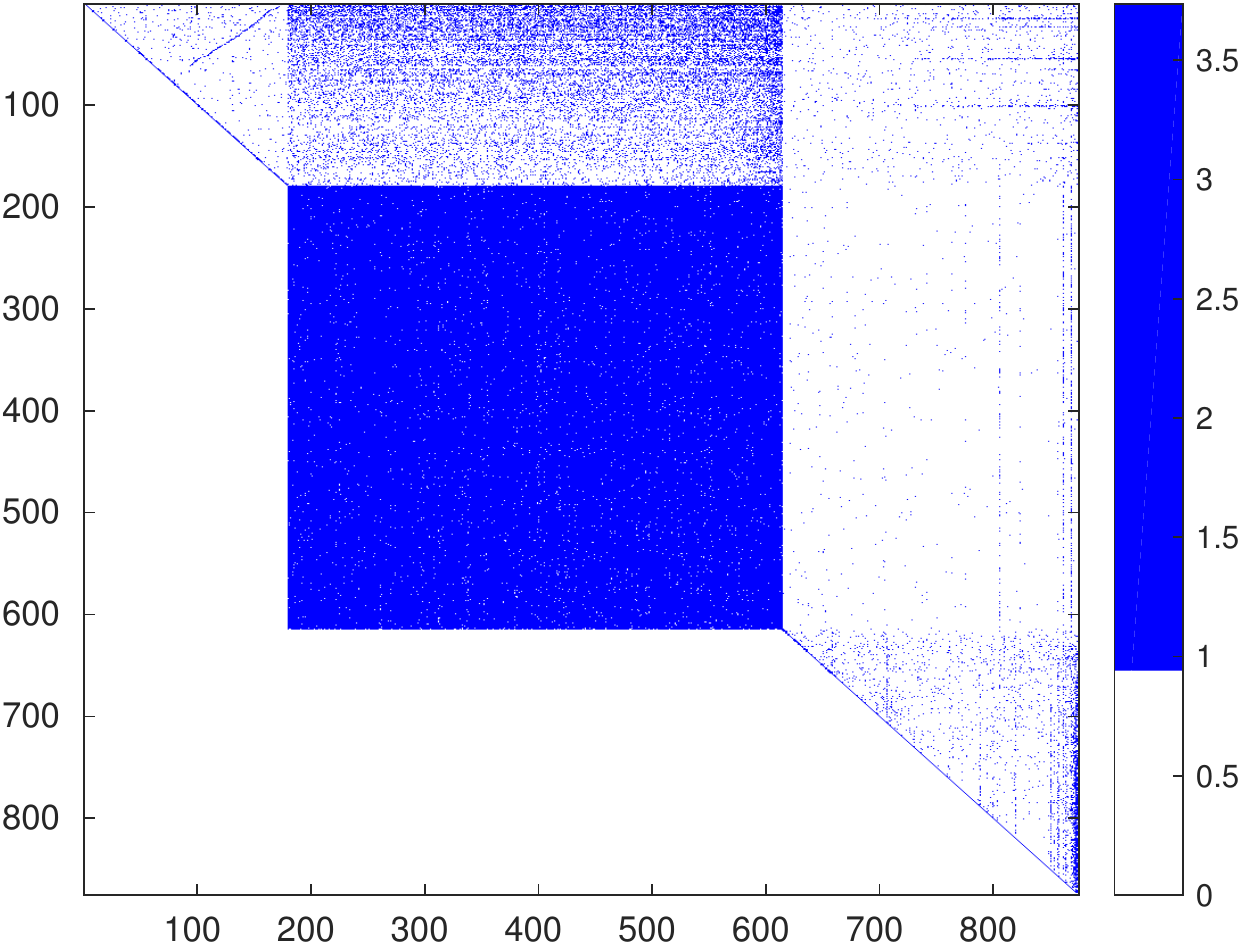}
  }
  \caption{
   \footnotesize The 1:1000 sparsity map of adjacency matrix of Google link graph \cite{google-net} with 875,713
    page nodes in {\bf (a)} the provided ordering and {\bf (b)} the ordering rendered by the DM decomposition, depicted by {\tt imagesc} in {\tt matlab}. Each point shows the number of non-zeros, in log scale, in the corresponding $1000 \times 1000$ block. The subnetwork in the middle of (b) is strongly connected with 434,818 nodes.}
  \label{fig:dm}
\end{figure}

%
Information networks in real world applications are not necessarily
irreducible and aperiodic as assumed by many existing iterative
solutions for guaranteed convergence. To meet such convergence
conditions, some heuristics were used to perturb or twist the network
structure with artificially introduced links~\cite{lee2003fast,
langville2006reordering}.
Instead, we decompose the network into strongly connected sub-networks
by applying the Dulmage-Mendelsohn (DM) decomposition algorithm
\cite{dulmage1958coverings} to the Laplacian matrix $I-P$. The DM
algorithm is highly efficient when diagonal elements are non-zero. 
It renders the matrix in block upper triangular form. See
Figure~\ref{fig:dm} for the Google link graph released by Google in 2002 \cite{google-net}.
Each diagonal block $B_{ii}$ corresponds to a subnetwork. A
square diagonal block corresponds to an irreducible
subnetwork. A non-zero off-diagonal block $B_{ij}$ in the upper part,
$i<j$, represents the links from cluster $j$ to cluster $i$. The top
block is associated with a {\em sink} cluster without outgoing links to
other cluster; the bottom block is associated with a {\em source}
cluster without incoming edges from other clusters. The solution for the
entire network can be obtained by the solutions to the subnetworks and
successive back substitution.
%

%


%

\subsection{Cascade of iterations} 
\label{subsec:cascade}
%

Several iterative methods exist for computing the PageRank vector by the
Brin-Page model. They include the power method by the eigenvector
equation (\ref{eqn:ppr-markov-equi}), and the Jacobi, Gauss-Seidel, and
SOR methods by equation (\ref{eqn:ppr-linear}). There are various
acceleration techniques used for calculating the PageRank vector, or a
small part of the vector, or even a single pair of nodes between
the personalized vector and the PageRank vector~\cite{kamvar2003extrapolation,
ilprints579, kamvar2004adaptive, DBLP:journals/corr/LofgrenBGS14}.
For the Brin-Page model, the iterative methods converge slower as
$\alpha$ increases and gets closer to $1$. We developed a cascading
initialization scheme. The solution to the model with $\alpha$ is used
as the initial guess to the iteration for the solution to the model with
$\alpha+\delta \alpha$, $\delta \alpha > 0$. Although it has accelerated
the computation with successively increased $\alpha$ values, this
technique is limited to sequential computation and ad hoc to the
Brin-Page model. We introduce next a novel algorithm without these
limitations.

 
%

\subsection{Shared invariant Krylov space} 
\label{subsec:SIK-space}
%

Our new algorithm for batch calculation of PageRank vectors with
multiple models and parameter values is based on the very fact that the
solutions to the models in Section~\ref{sec:pageRank-models} all reside
in the same Krylov space,
\begin{equation} 
\label{eqn:Krylov-space} 
{\cal K}(P, v) = \{v, Pv, P^2v, \cdots, P^{k}v, \cdots  \}, 
\quad 
\begin{array}{l} v\geq 0 \\ e^{\rm T}v = 1 \end{array}. 
\end{equation} 
The space has the property $P {\cal K}(P,v) = {\cal K}(P,v)$, i.e., it
is a spectrally invariant subspace. In PageRank terminology,
${\cal K}(P,v)$ is a personalized invariant subspace.  
We have the remarkable fact about the model family in Section~\ref{sec:pageRank-models}. 
\newtheorem{theorem}{Theorem}
\begin{theorem}
\label{thm:krylove-functions}
Any model solution (\ref{eqn:uni-random-walk}), at any particular
damping parameter value, and its
trajectory (\ref{eqn:trajectory-equation}) are functions in the Krylov
space ${\cal K}(P,v)$.
\end{theorem}

\begin{theorem}[]
\label{thm:krylov}
Let $m=\mbox{dimension}({\cal K}) $.  Denote by $K$ the matrix
composed of the Krylov vectors.  Let $K = QR$ be the QR factorization of
$K$.  Then, $Qe_1 =v$ and $PQ = QH$, where $H$ is an $m\times m$ upper
Hessenberg matrix, and $e_1$ is the first column of the identity matrix $I$.
\end{theorem}

A few remarks. Matrix $H$ in Theorem~\ref{thm:krylov} is the
representation of matrix $P$ under basis $Q$ in the Krylov space.  In
numerical computation, we use a rank-revealing version of the QR
factorization with $Qe_1 = v$.
In theory, the dimension $m$ is equal to the number of spectrally
invariant components of $P$ that present in $v$.  In the extreme case,
$m=1$ when $v$ is the Perron vector of $P$. In general, by the condition
$e^{\rm T}v =1$, $v$ is not deficient in Perror component. In our study
on real-world graphs, which we will detail shortly in Section~\ref{sec:numerical-experiments},
the numerical dimension is low, matrix $H$ is therefore small.  We may
view this as a manifest of the smallness of the real-world graphs under
study.  We exploit these theoretical and practical facts.

\newtheorem{corollary}{Corollary}
\setcounter{corollary}{2}
\begin{corollary}[]
\label{col:krylov}
For any function $g$ in the Krylov space (\ref{eqn:Krylov-space}), we
have $g(P)v = Qg(H)e_1$.
\end{corollary}
When dimension $m$ is modest, we translate by Corollary \ref{col:krylov}
the calculation of $x_{g}=g(P)v$ with $N\times N$ matrix $P$ on vectors
to the calculation of $ \hat{x}_{g} = g(H)e_1$ with $m\times m$ matrix
$H$ on vectors, followed by a matrix-vector product $Q\hat{x}_g$. The
vector $\hat{x}_g$ is the spectral representation of $x_g$ in the Krylov
space.  In the model family, solutions $x_f$ (\ref{eqn:uni-random-walk})
differ from one to another in their spectral representations
$\hat{x}_f$, they share the same basis matrix $Q$ in the ambient network
space.

Our algorithm consists of the following major steps. Let
${\cal G}= \{ g\}$ be a set of functions under study.
\begin{inparaenum}[(1)] 
\item Calculuate Krylov vectors to form matrix $K$ in
  Theorem~\ref{thm:krylov}, apply rank-revealing $QR$ factorization to
  $K$, and find numerical dimension $m$; \footnote{These substeps are
    integrated in pratical computation in order to determine quickly a
    sufficient number of Krylov vectors.}
\item Construct the matrix $H$, by Theorem~\ref{thm:krylov},
  from $R$ and the permutation matrix $\Pi$ rendered by the
  rank-revealing $QR$;
\item Calculate $\hat{x}_f= g(H)e_1$ for all functions in ${\cal G}$; 
\item Transform $\hat{x}_f$ from the Krylov-spectral space to the
  ambient network space by the same basis matrix $Q$, based on
  Corollary~\ref{col:krylov}.
\end{inparaenum}

 
%

 
%


\section{Experiments on real-world link graphs} 
\label{sec:numerical-experiments}
%
%

We show in numerical values how PageRank vector responses to variation
in damping variable with each model and across models, on $6$ real-world
link graphs.

\subsection{Experiment setup: data and models}
\label{subsec:data-and-models} 
%

The $6$ link graphs we used for our experiments are publicly available
at {\em the Koblenz Network Collection}\cite{kunegis2013konect}. 
The basic information of the graphs is summarized in
Table~\ref{tab:dataset}, where $\max(d_{out})$ is the maximum out-degree
(the number of citations) of graph nodes, $\max(d_{in})$ is the maximum 
in-degree (the number of backlinks), $\mu(d_{out}) = \mu(d_{in})$ is
the average out-degree, which equals to the average in-degree, 
and LSCC stands for the largest strongly connected component(s) of the graph. 
The Google graph of today is reportedly containing hundreds of trillions 
of nodes, substantially larger than the snapshot size used here.
%
%

\begin{table}[H]
\centering
\caption{Dataset Description}
\label{tab:dataset}
\begin{adjustbox}{width=0.5\textwidth}
\begin{tabular}[t]{lrrr}
\toprule
&  Total \#nodes & \#nodes in LSCC & $\left[\max(d_{out}), \mu(d_{out}), \max(d_{in})\right]$\\
\midrule
   Google \cite{google-net} & 875,713 & 434,818 & $[ 4209, 8.86, 382]$ 
\\ 
Wikilink \cite{kunegis2013konect} & 12,150,976 & 7,283,915 & $[7527, 50.48, 920207]$ 
\\
DBpedia \cite{auer2007dbpedia} & 18,268,992 & 3,796,073 & $[8104, 26.76, 414924]$ 
\\
Twitter(www)\cite{Kwak10www} & 41,652,230 & 33,479,734 & $[2936232, 42.65, 768552]$  
\\
Twitter(mpi)\cite{icwsm10cha} & 52,579,682 & 40,012,384 & $[778191, 47.57, 3438929]$
\\
friendster\cite{yang2015defining} & 68,349,466 & 48,928,140 & $[3124, 32.76, 3124]$  
\\ 
\bottomrule{}
\end{tabular}
\end{adjustbox}
\end{table}


%
%
For variation analysis of each graph in Table~\ref{tab:dataset}, the 
associated link matrix $P$ is well specified. We use the same personalized
or customized distribution vector $v$, which we get by drawing elements from 
standard Gaussian distribution ${\cal N}(0, 1)$, followed by 
normalization $v^{\rm T}e = 1$.
We report variation analysis results with three particular models :
Brin-Page model (\ref{eqn:ppr-linear}) with damping variable $\alpha$,
Chung's model (\ref{eqn:heat-model}) with variable $\beta$, and the 
log-$\gamma$ model (\ref{eqn:log-gamma}). The last is used as an 
illustration of new models in the family (\ref{eqn:uni-random-walk}).

%

\subsection{Variation in PageRank vector} 
\label{subsec:intro-model-analysis}
%
%


\begin{figure}[!htb]
  \centering
  \subfigure[$x(\rho)$ of Google link graph]{
    \includegraphics[width=0.22\textwidth]
        {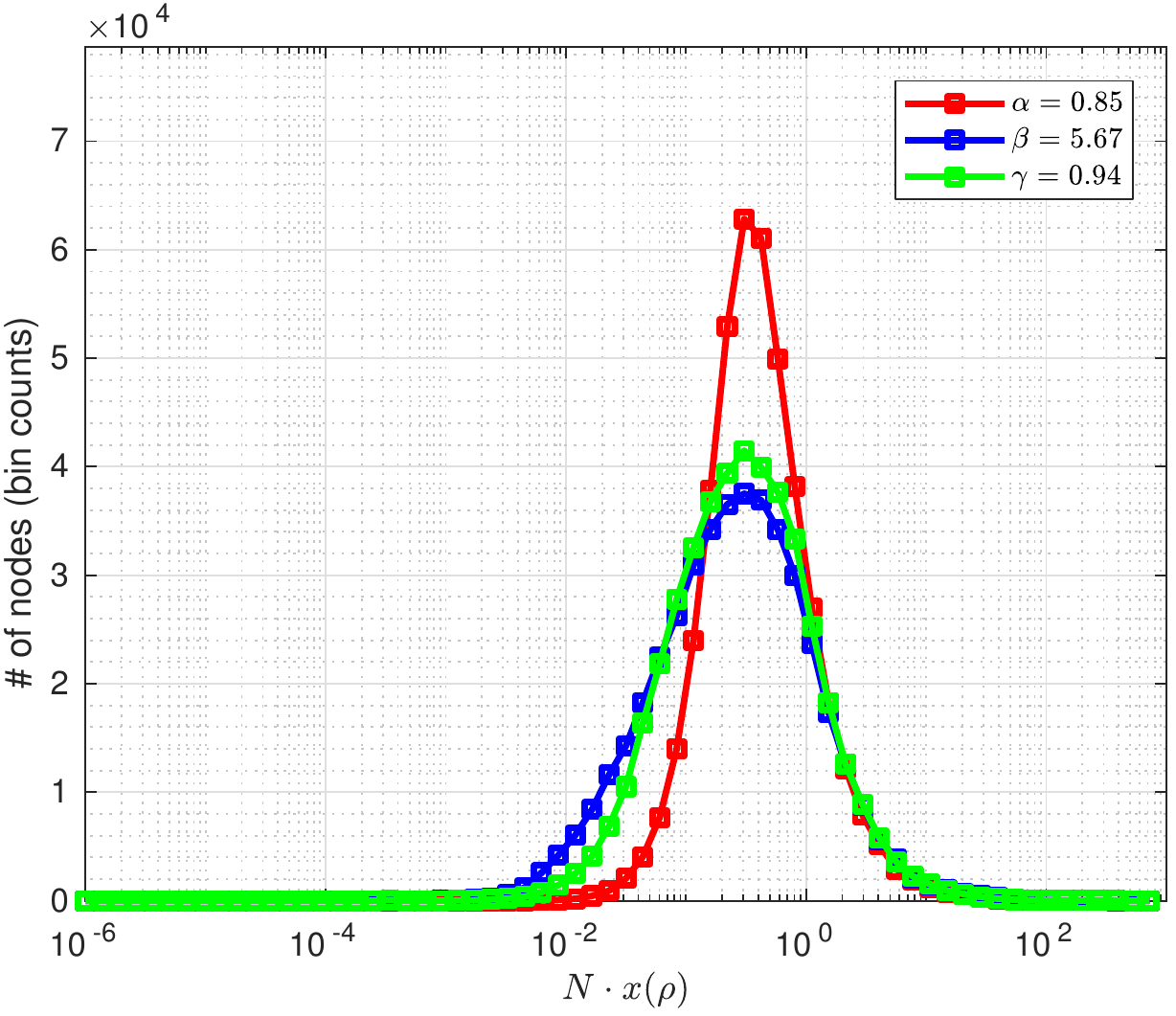}
    \label{fig:google-hist-unify}
  }
  \subfigure[$x(\rho)$ of Twitter(www) link graph]{
    \includegraphics[width=0.22\textwidth]
       {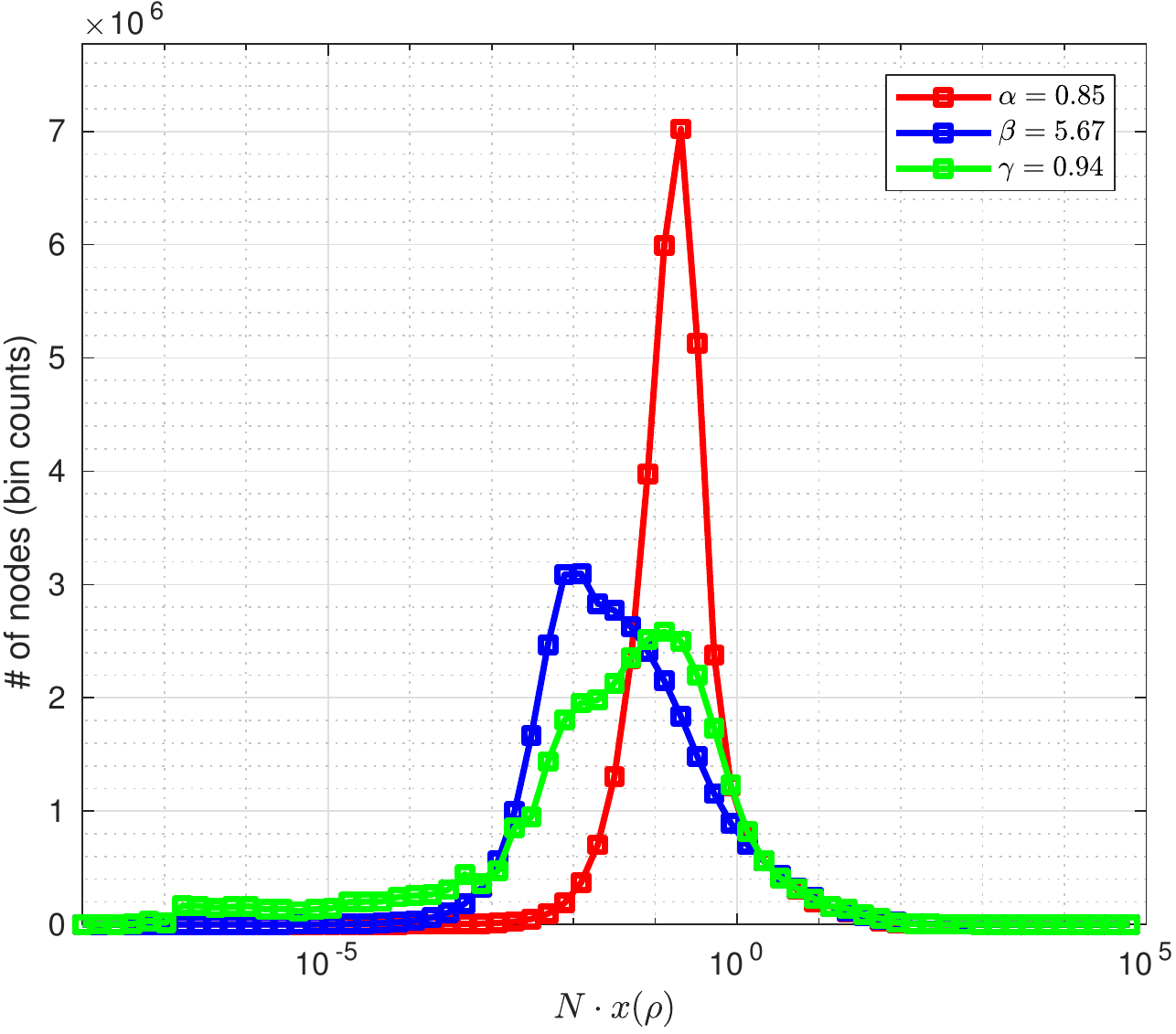}
    \label{fig:twitter-hist-unify}
  }
  \caption{\footnotesize%
    Inter-model comparison of histograms of $N\cdot x_f(\rho)$ among the 
    three models with corresponding parameter values $\alpha_{o}$,
    $\beta_{o}$ and $\gamma_{o}$ so that the expected value of
    walking steps for each model is
    $ \alpha_{o}/(1-\alpha_{o})= 5.\dot{6} $ with $\alpha_{o} = 0.85$,
    see the model correspondence equalities (\ref{eqn:param-corres}).
    (a) comparison on Google link graph; (b) comparison on Twitter(www)
    link graph. }
\label{fig:unify-model-google}
\end{figure}


%
%
In order to show the quantitative response in PageRank vector $x_f$
over $N$ nodes with the variation in damping variable $\rho$, we display
the histogram of $N\cdot x_f(\rho)$ for model $f$ at parameter value
$\rho$. In Figure~\ref{fig:unify-model-google} we show the
histograms associated with three models on Google network. The parameter for Brin-Page
model is set to the value $\alpha_{o} = 0.85$. The parameter for 
the other two models are set by (\ref{eqn:param-corres}).
%
We observe that the histogram with Brin-Page model has higher and narrower peaks
than Chung's model. The histogram of
log-$\gamma$ model is in between. This is expected by the relationships
in the damping weights among the three models, as discussed in
Section~\ref{subsec:damping-corres}.

%

\begin{figure}[!htb]
  \centering
  \subfigure[Brin-Page model]{
    \includegraphics[width=0.14\textwidth]
     {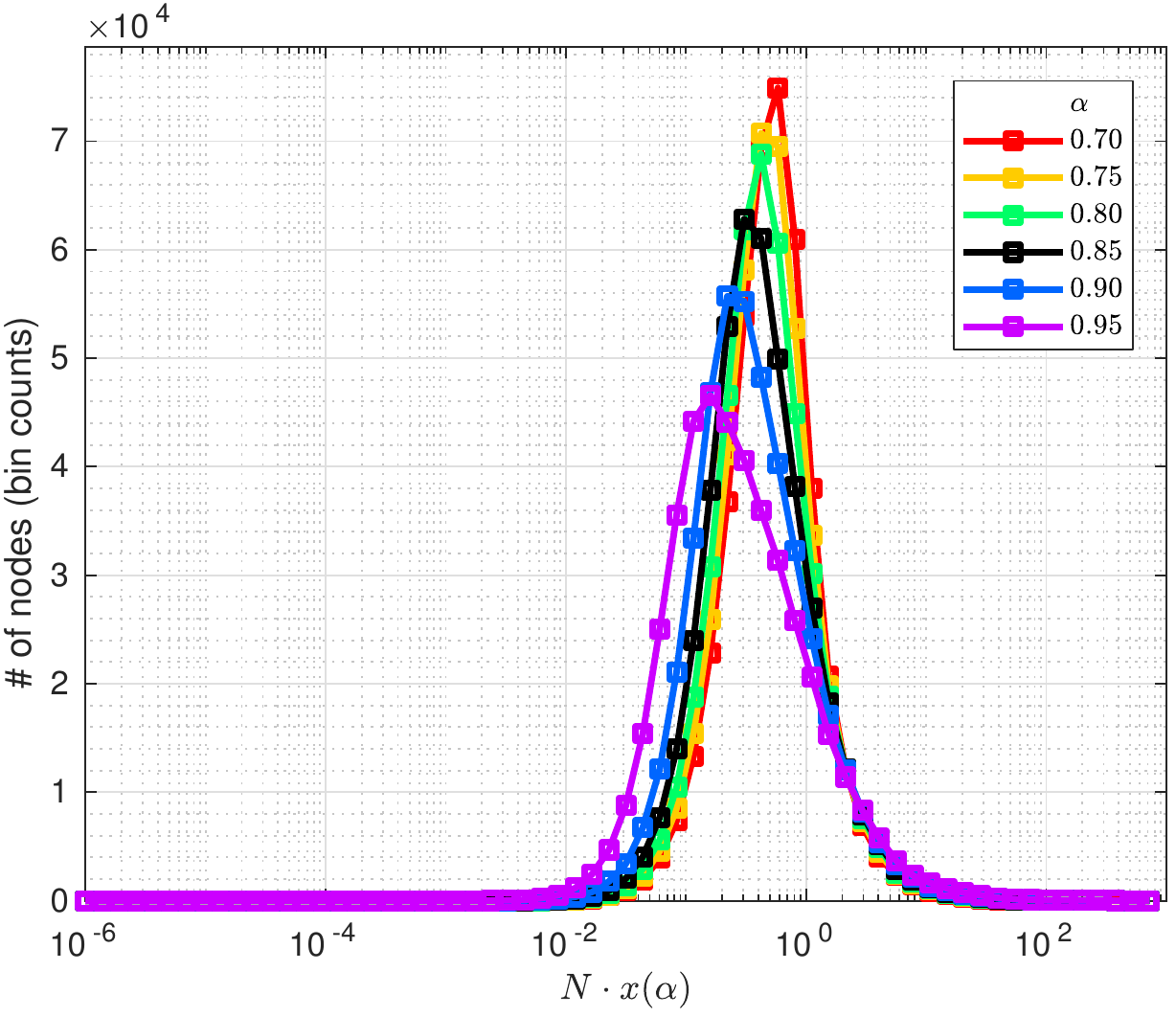}
    \label{fig:google-hist-bp}
  }
  \subfigure[log-$\gamma$ model]{
    \includegraphics[width=0.14\textwidth]
    {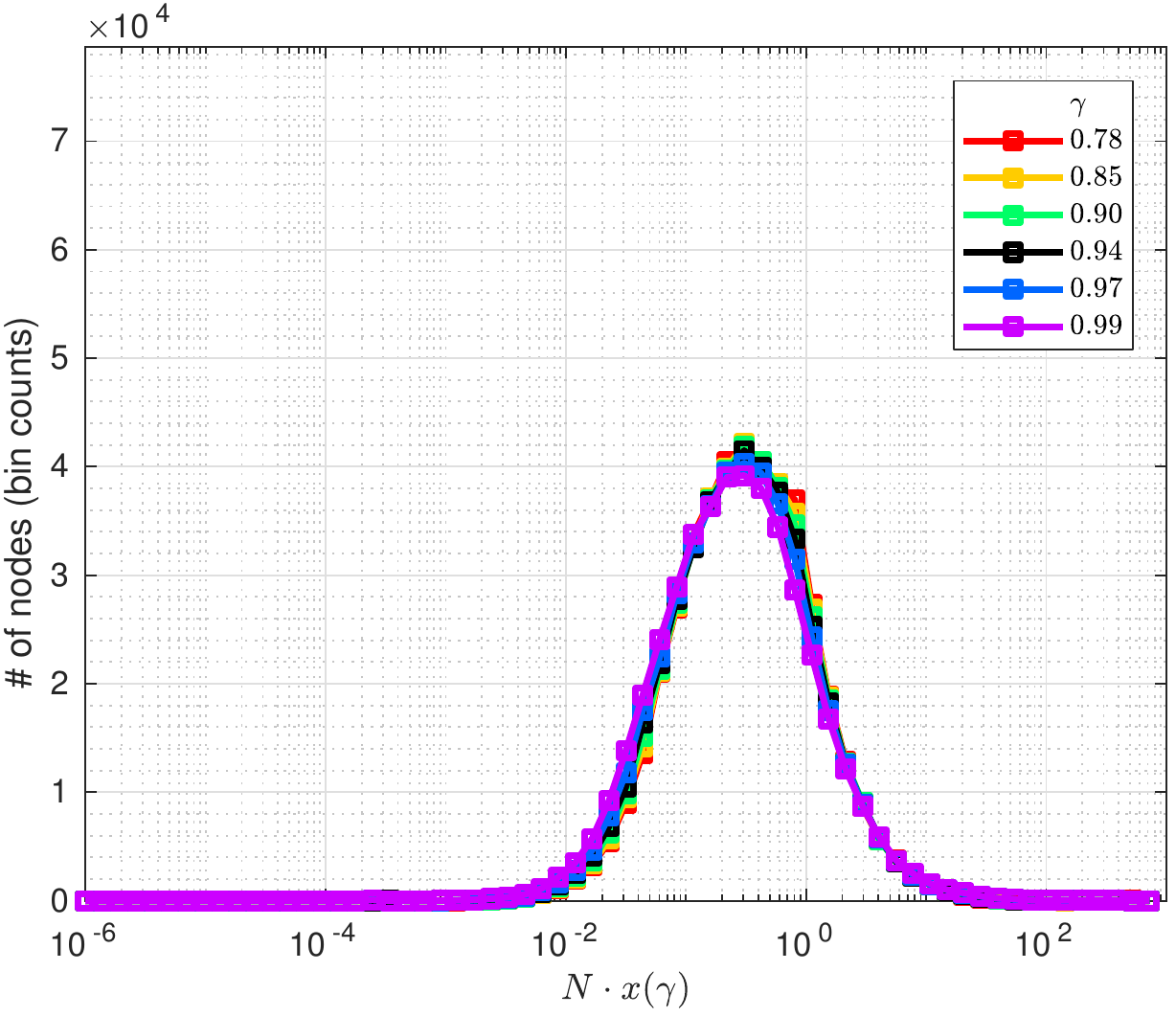}
    \label{fig:google-hist-chung}
  }
  \subfigure[Chung's model]{
    \includegraphics[width=0.14\textwidth]
    {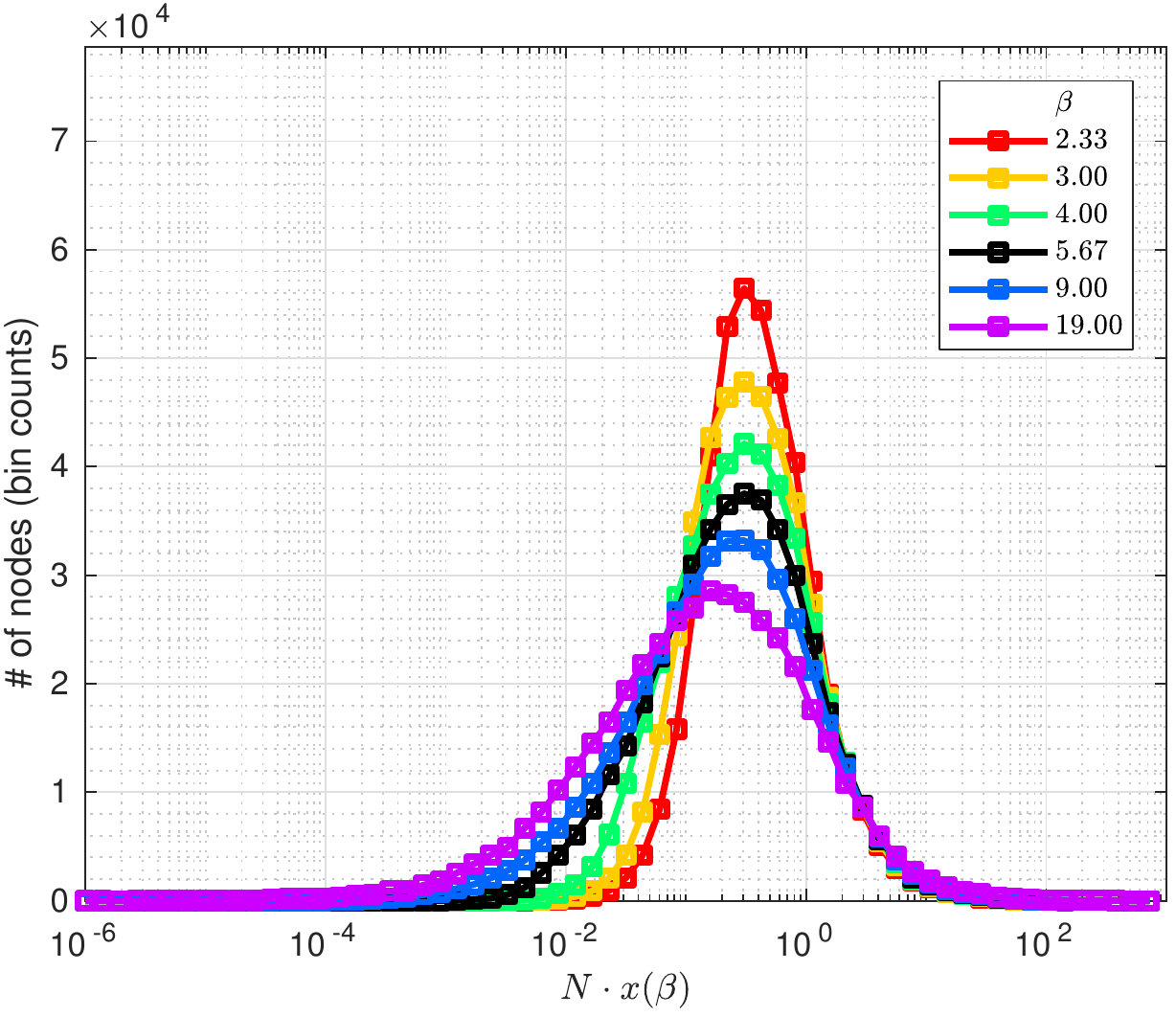}
    \label{fig:google-hist-chung}
  }
  \subfigure[Brin-Page model]{
    \includegraphics[width=0.14\textwidth]
    {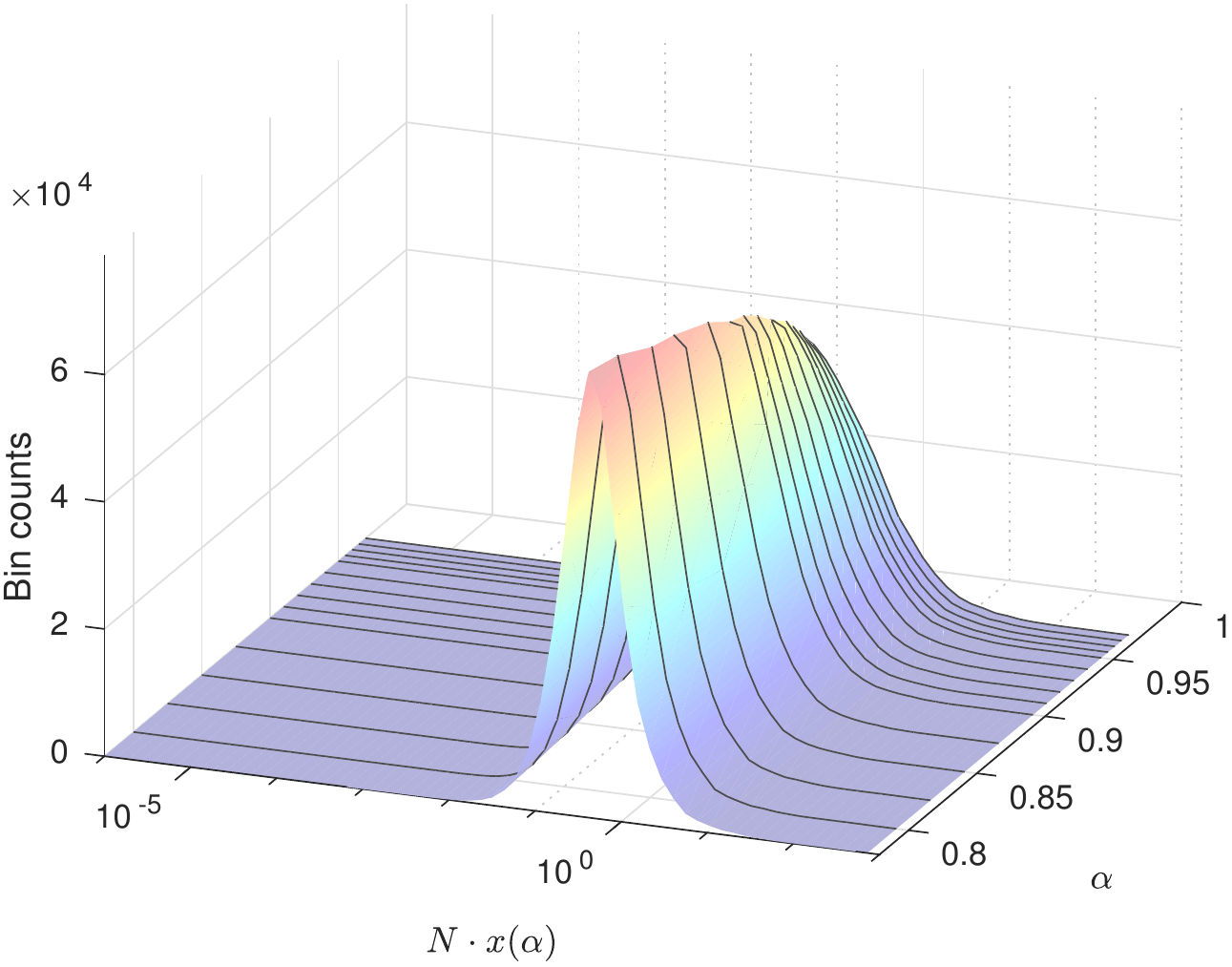}
    \label{fig:google-hist-bp-3d}
  }
  \subfigure[log-$\gamma$ model]{
    \includegraphics[width=0.14\textwidth]
    {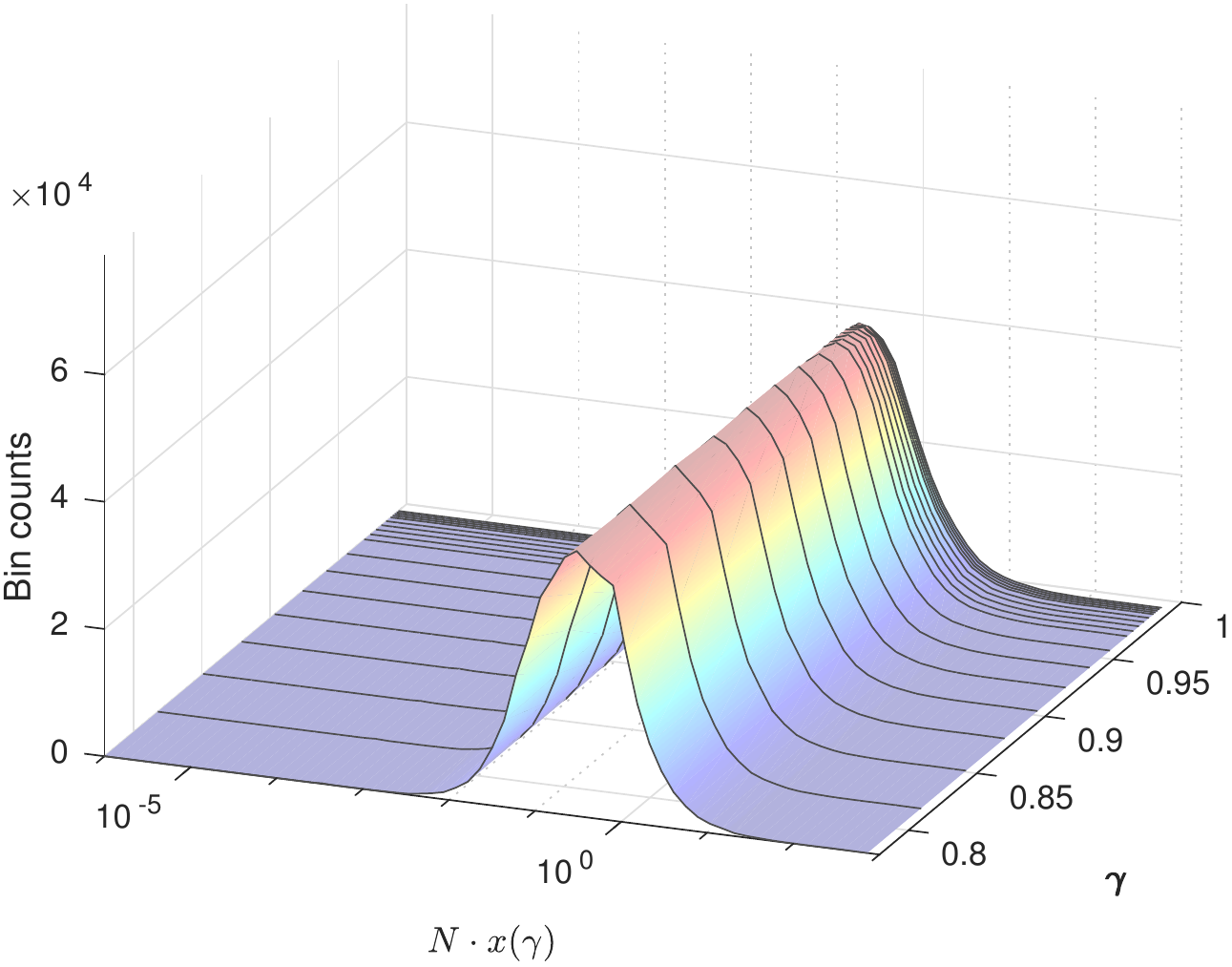}
    \label{fig:google-hist-chung-3d}
  }
  \subfigure[Chung's model]{
    \includegraphics[width=0.14\textwidth]
    {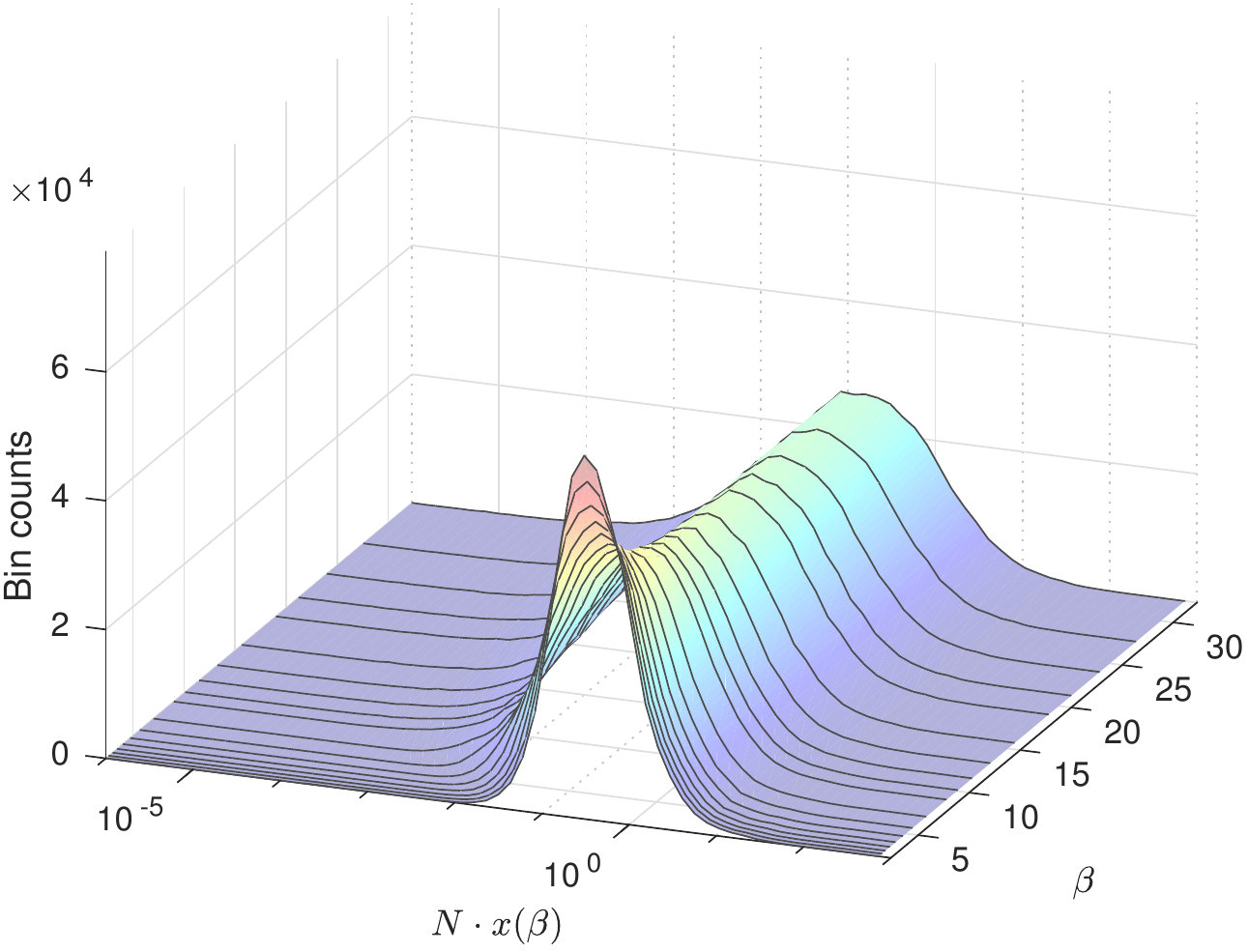}
    \label{fig:google-hist-chung-3d}
  }

  \caption{\footnotesize 
    Comparison in histograms of $N\cdot x_f(\rho)$ on the Google link
    graph over a range of damping variable value. 
    {\bf Left column:} Brin-Page model, 
    {\bf Middle column:} log-$\gamma$ model, 
    {\bf Right column:} Chung's model.
    {\bf Top row:} 2D display of 6 histograms associated with
    6 parameter values shown in the respective legends.  The
    histogram in black with Brin-Page model is associated with the value
    $\alpha = 0.85$.  The corresponding parameter values with Chung's
    model and log-$\gamma$ model are set by (\ref{eqn:param-corres}), 
    the associated histograms are color
    coded by the corresponding parameter values.
    {\bf Bottom row:} a stack of multiple histograms shown in 3D space over
    the range $\alpha \in [0.7, 0.97]$ with Brin-Page model,
    $\beta \in [2.\dot{6}, 32.\dot{3}] $ with Chung's model, and $\gamma \in [0.7787,0.994]$ with log-$\gamma$ model. The histograms with Chung's
    model have flattened peaks at larger values (toward the back
    end). The log-$\gamma$ model is nearly insensitive to $\gamma$ change in the range above.}  %
\label{fig:histogram-google}
\end{figure}


%
%

\begin{figure}[!htb]
  \centering
  \subfigure[Brin-Page model]{
    \includegraphics[width=0.14\textwidth]
        {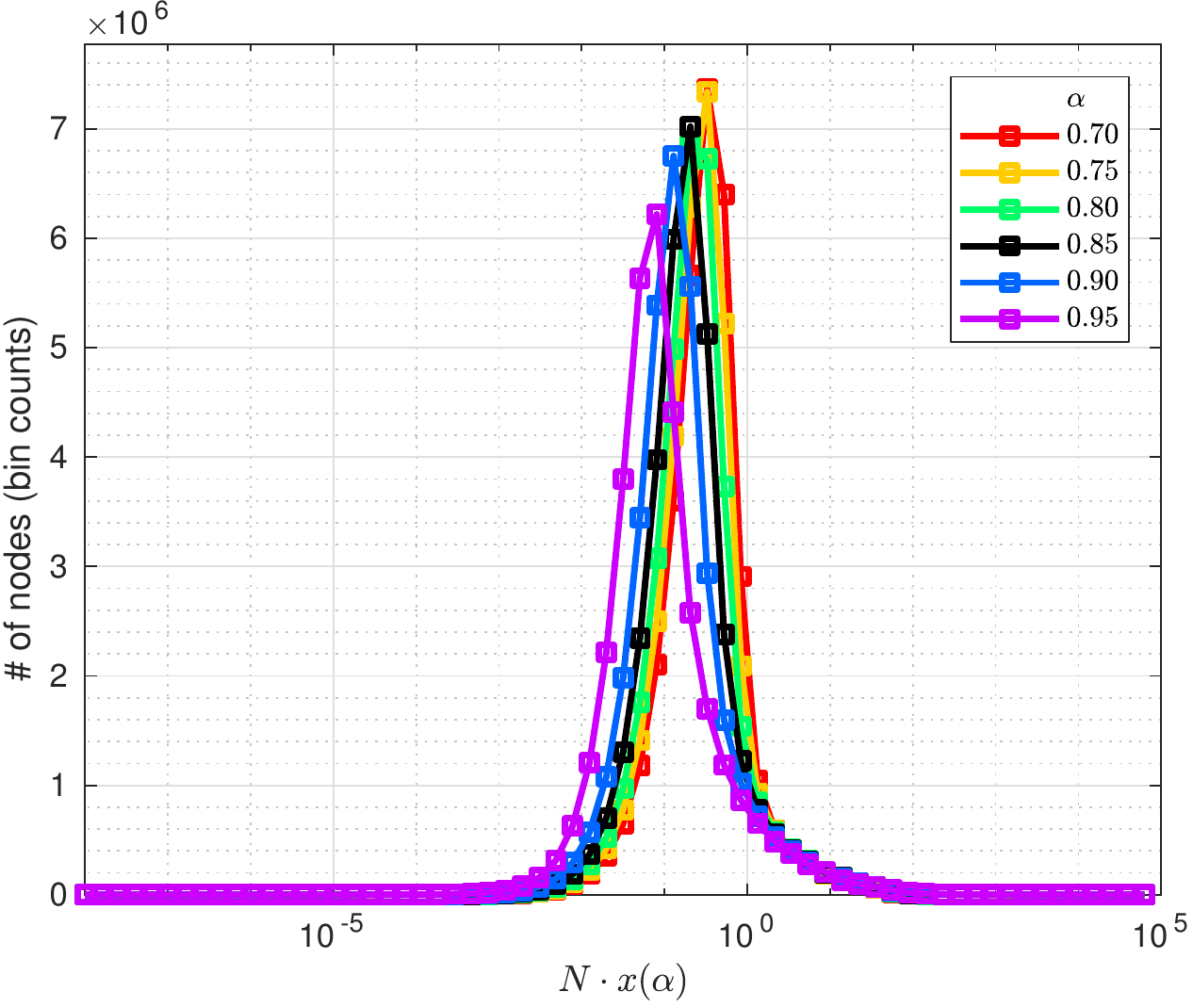}
    \label{fig:twitter-hist-bp}
  }
  \subfigure[log-$\gamma$ model]{
    \includegraphics[width=0.14\textwidth]
    {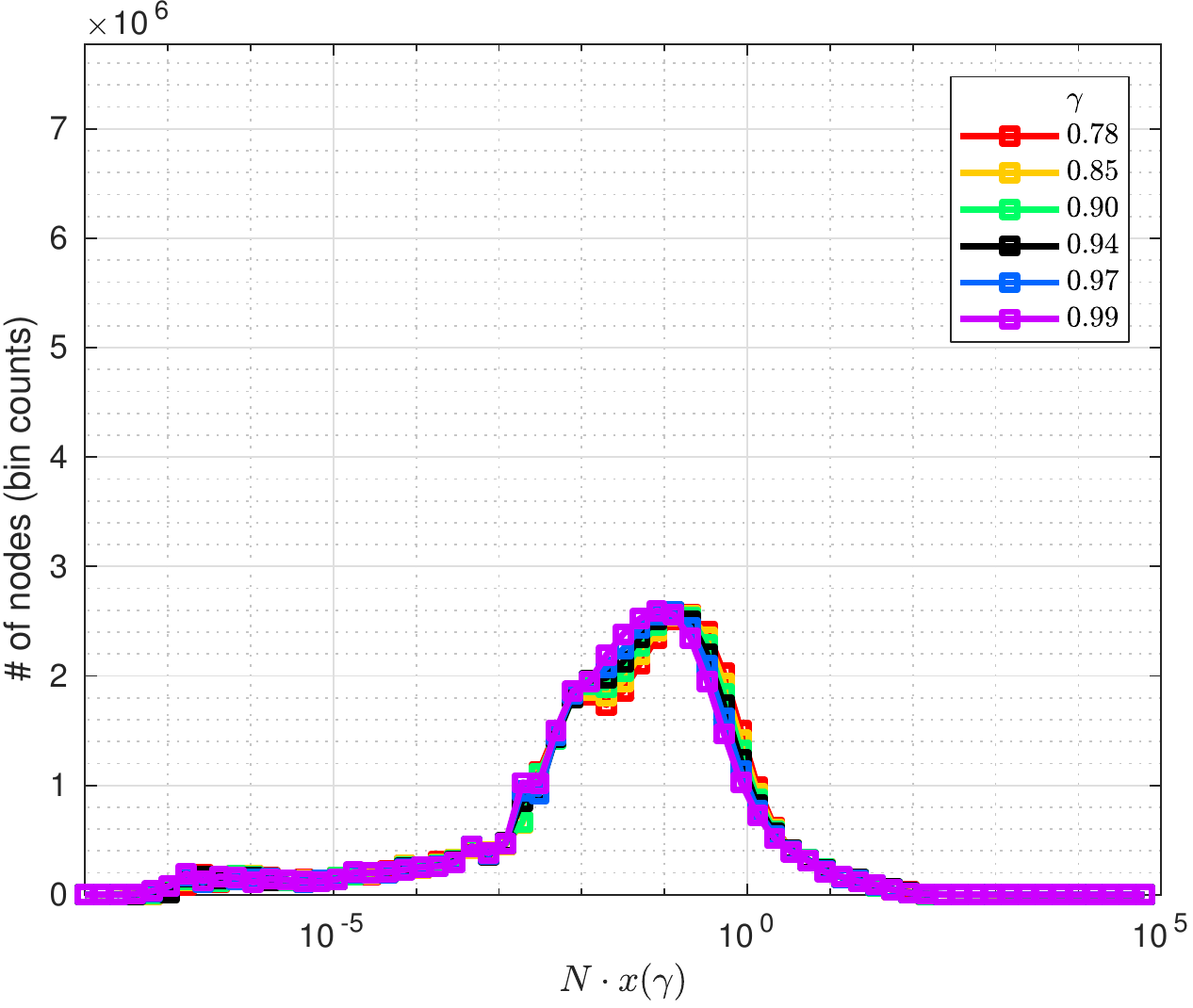}
    \label{fig:google-hist-chung}
  }
  \subfigure[Chung's model]{
    \includegraphics[width=0.14\textwidth]
        {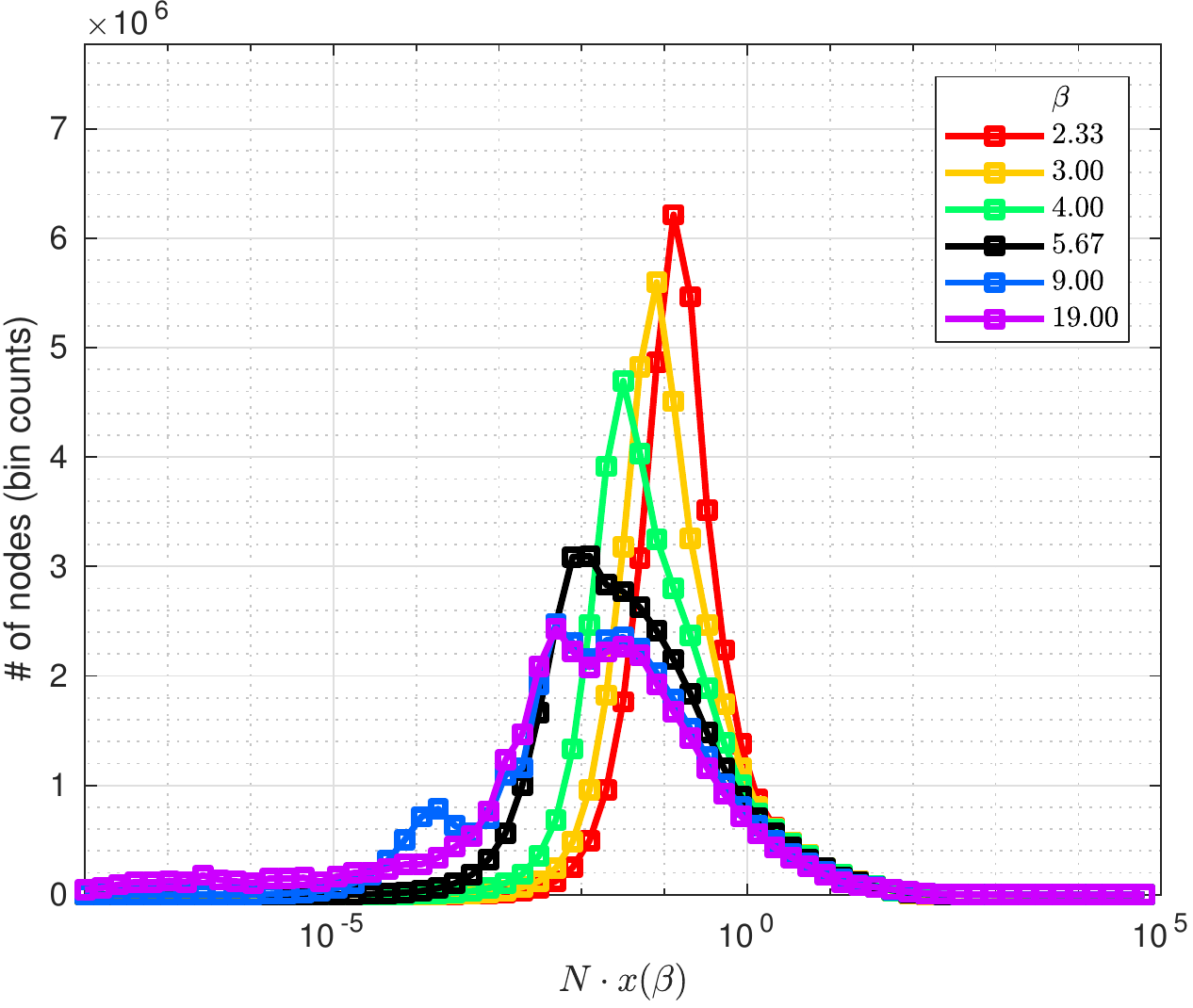}
    \label{fig:twitter-hist-chung}
  }
  \subfigure[Brin-Page model]{
    \includegraphics[width=0.14\textwidth]
        {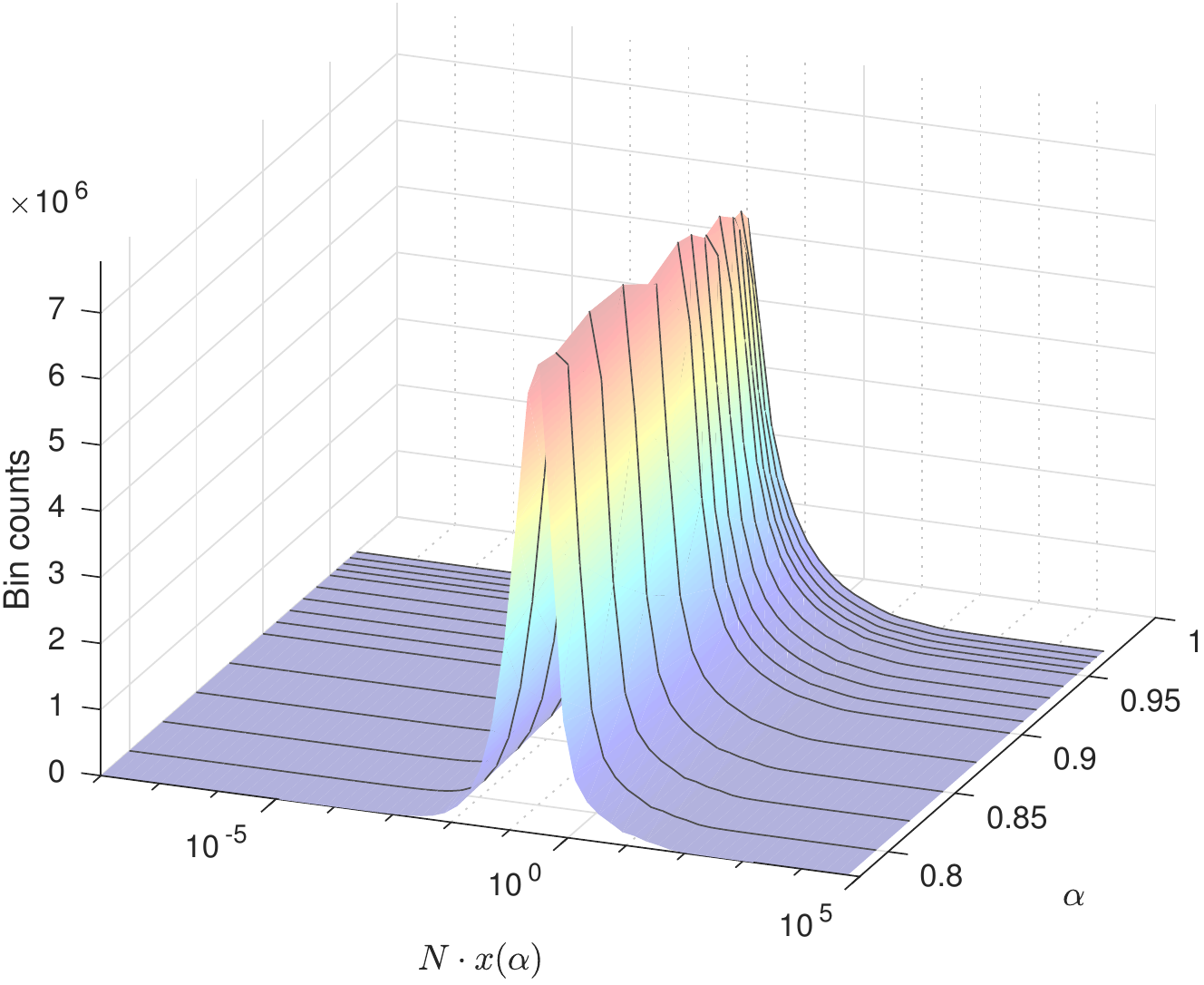}
    \label{fig:twitter-hist-bp-3d}
  }
  \subfigure[log-$\gamma$ model]{
    \includegraphics[width=0.14\textwidth]
    {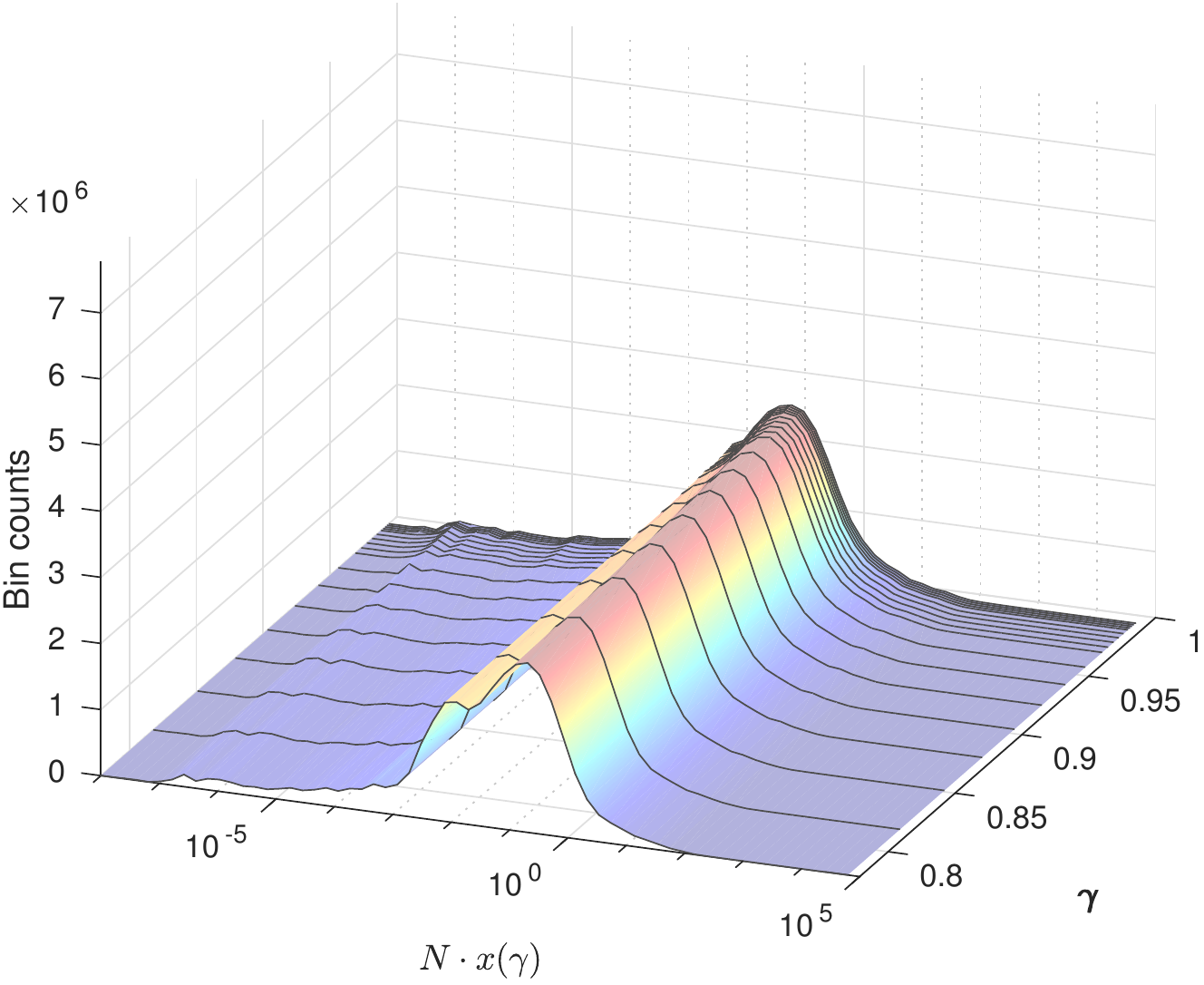}
    \label{fig:google-hist-chung-3d}
  }
  \subfigure[Chung's model]{
    \includegraphics[width=0.14\textwidth]
        {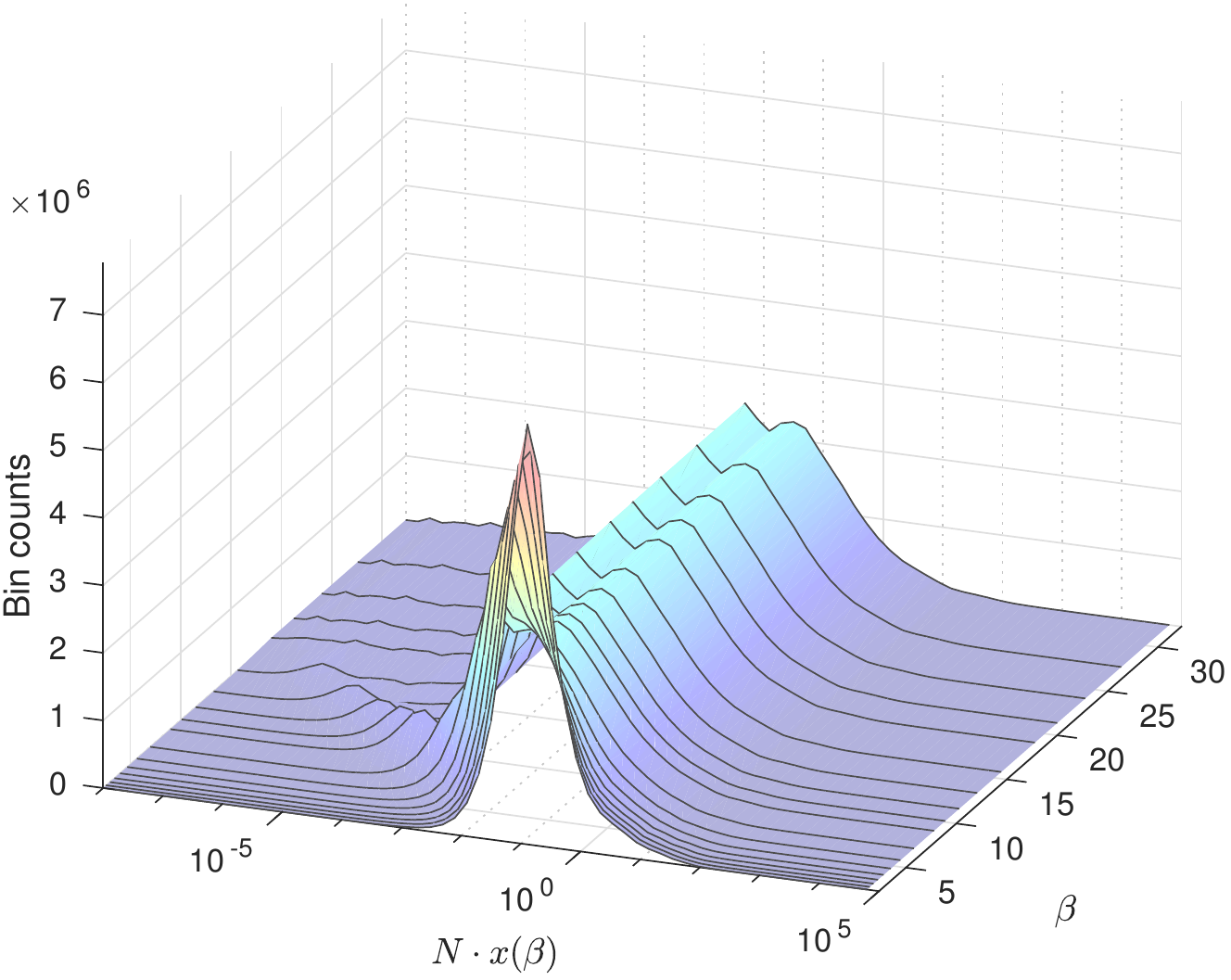}
    \label{fig:twitter-hist-chung-3d}
  }

  \caption{\footnotesize 
  Comparison in histogram of $N\cdot x_f(\rho)$ on the Twitter graph 
  over a range of dampling value, in the same settings as in
  Figure~\ref{fig:histogram-google}.
  }
\label{fig:histogram-twitter}
\end{figure}


%
%
Figure \ref{fig:histogram-google} and Figure \ref{fig:histogram-twitter} show the variation in the histograms over a range of the damping
variable per model as well as the comparison side by side between
the three models on two datasets. The models
have similar behaviors on the other $4$ graphs in Table~\ref{tab:dataset}.
Supplementary material can be found in \cite{MS-qian-2018}. With larger damping factors in the models, 
the distribution become less centralized. Log-$\gamma$ model, specifically, is less sensitive to $\gamma(\alpha)$
range with $\alpha \in [0.7,0.97]$.
%


%

\subsection{Relative variation measured by  KL divergence} 
\label{subsec:intro-model-analysis}
%


\begin{figure}[!htb]
  \centering
  \subfigure[{\footnotesize $\alpha_o=0.85$}]{
    \includegraphics[width=0.14\textwidth]
          {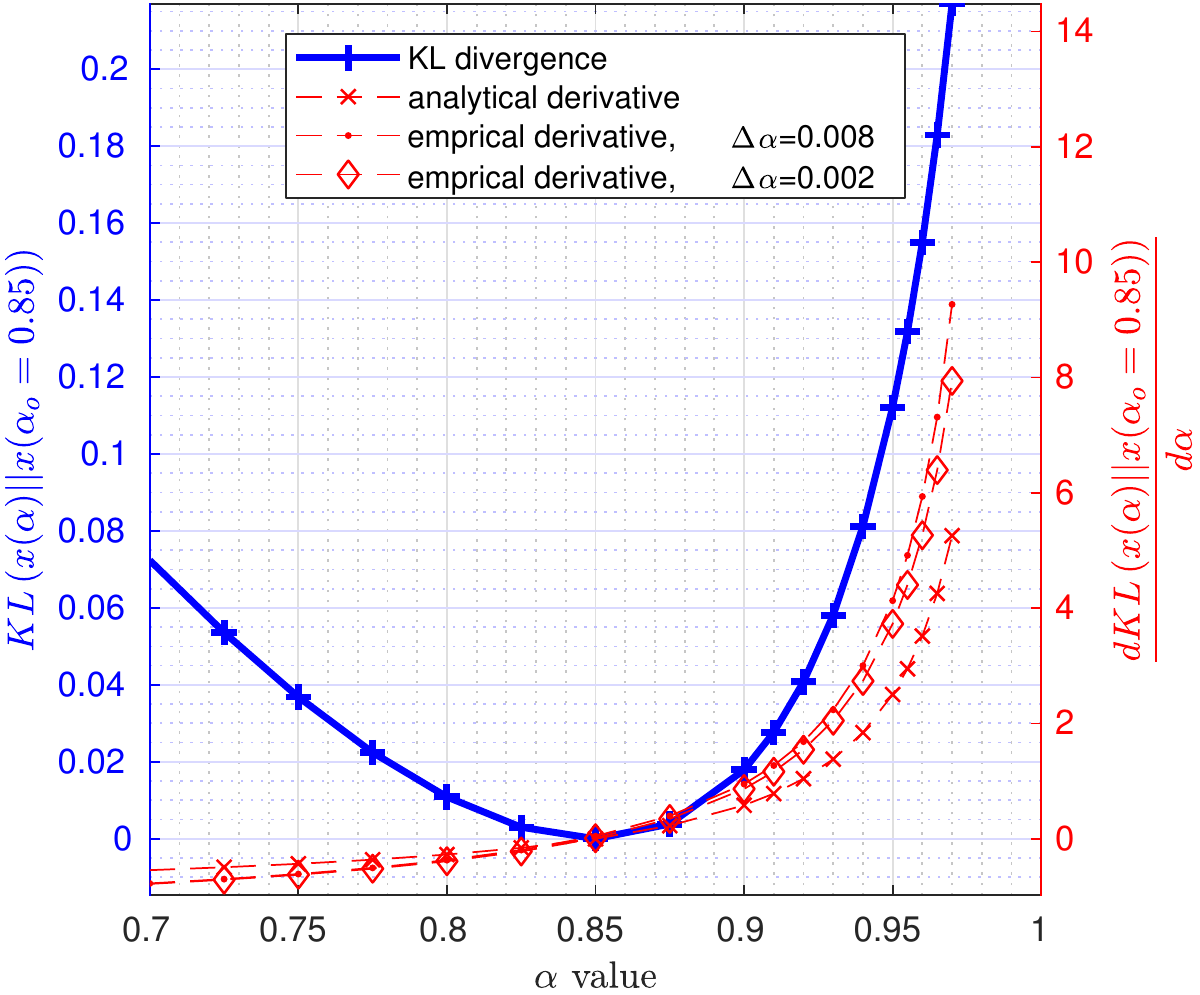}
  }
  \subfigure[{\footnotesize $\gamma_o=0.94146$}]{
    \includegraphics[width=0.14\textwidth]
          {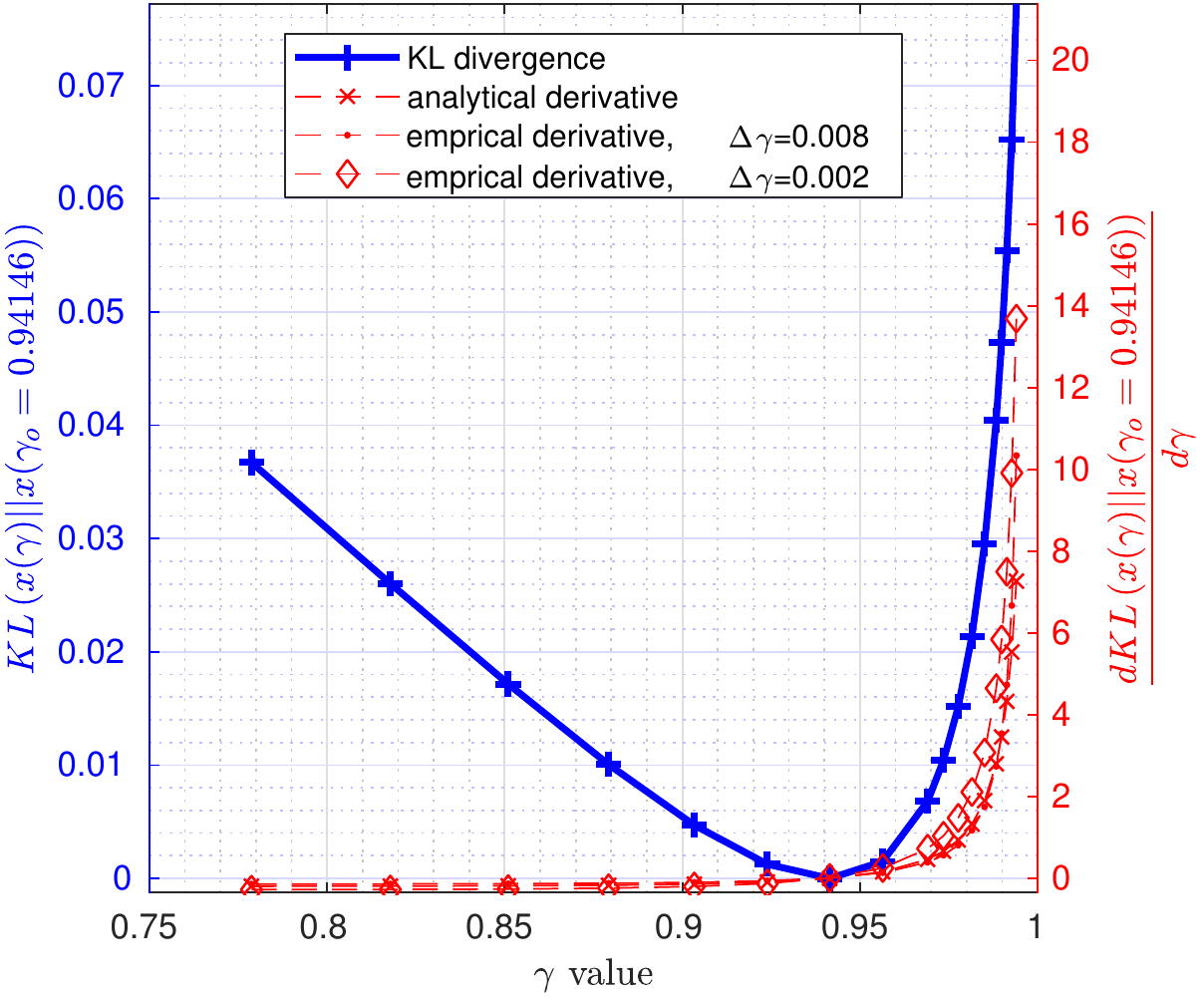}
  }
  \subfigure[{\footnotesize $\beta_o=5.\dot{6}$}]{
    \includegraphics[width=0.14\textwidth]
          {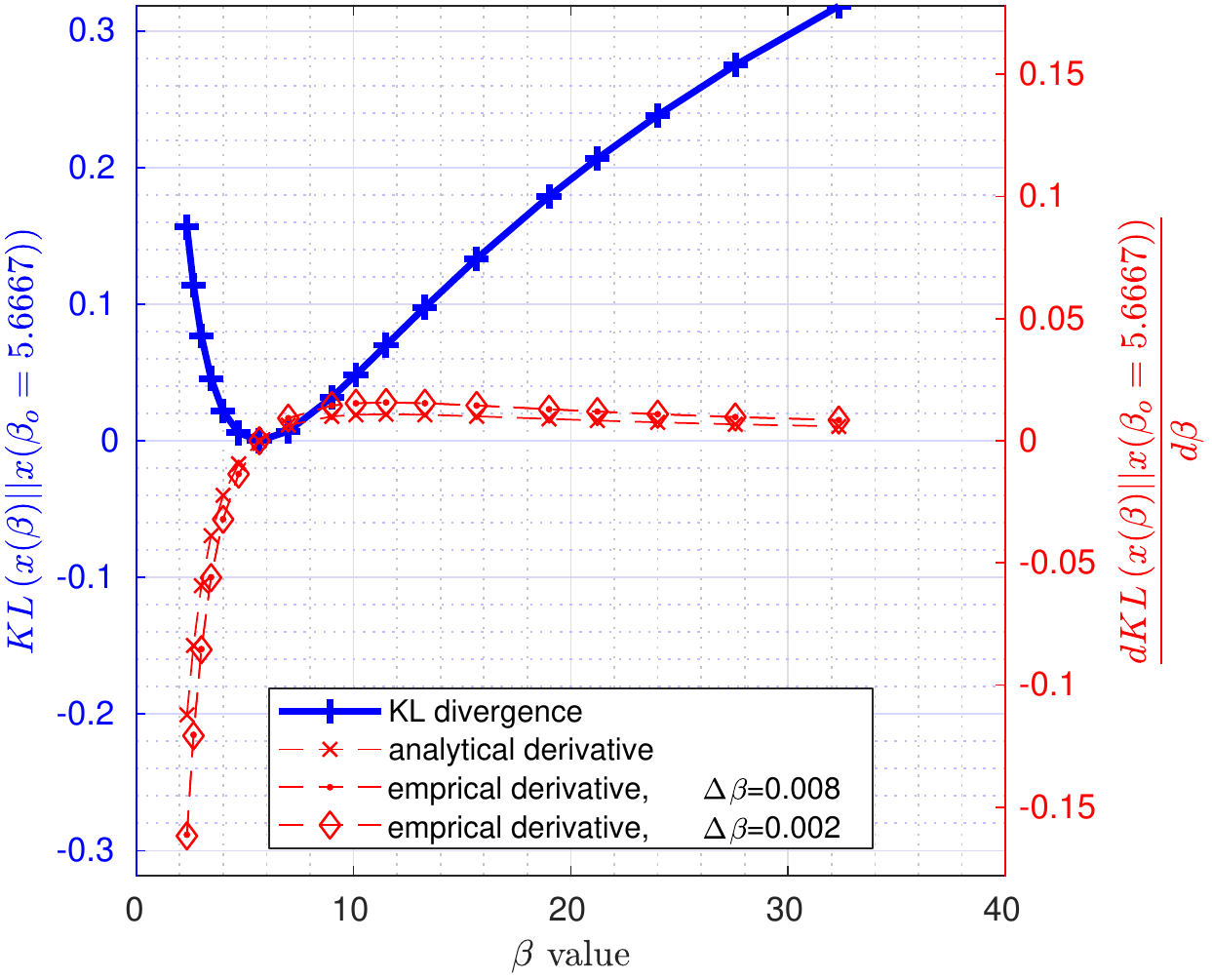}
  }
  \subfigure[{\footnotesize $\alpha_o=0.95$}]{
    \includegraphics[width=0.14\textwidth]
          {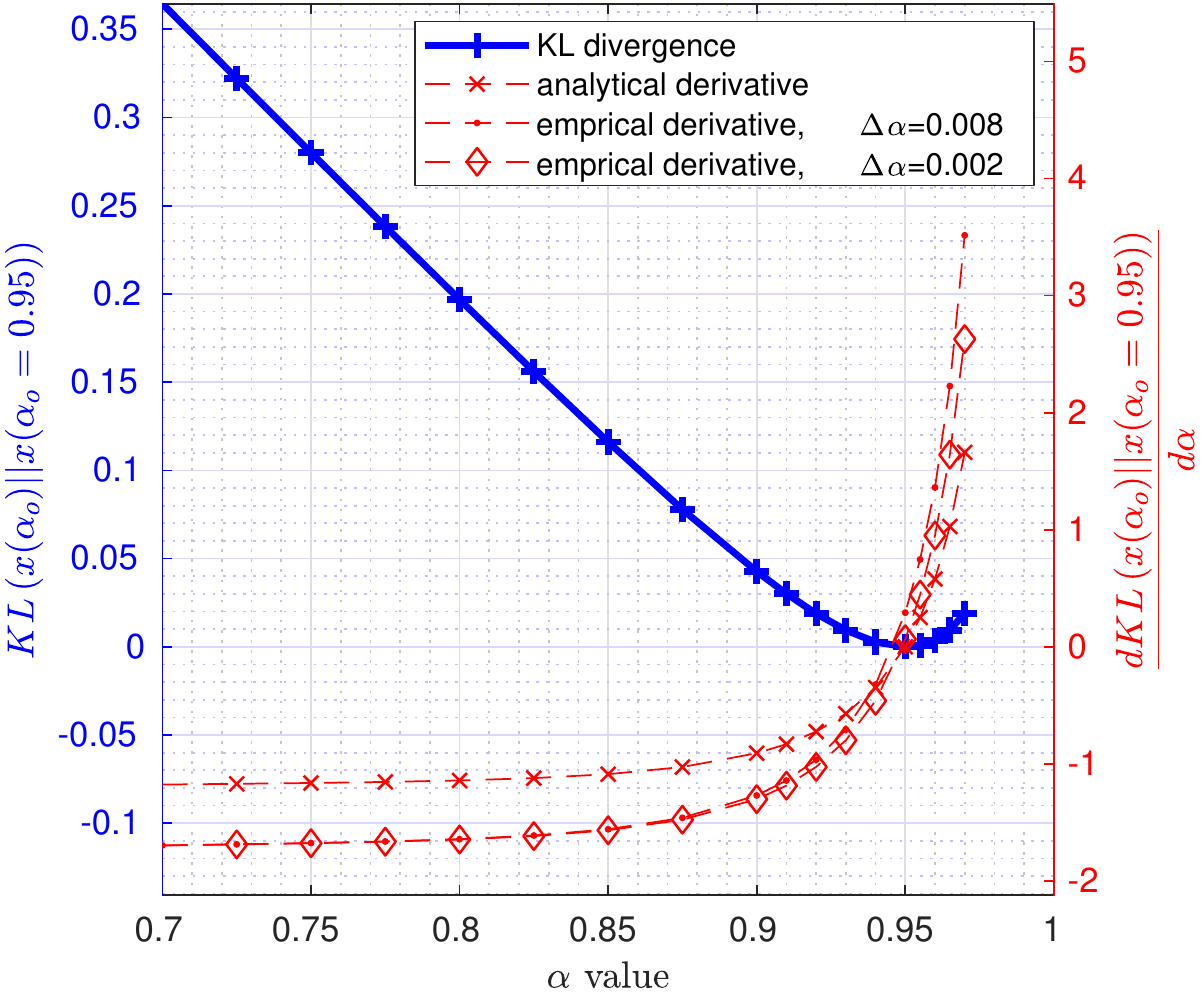}
  }
  \subfigure[{\footnotesize $\gamma_o=0.98831$}]{
    \includegraphics[width=0.14\textwidth]
          {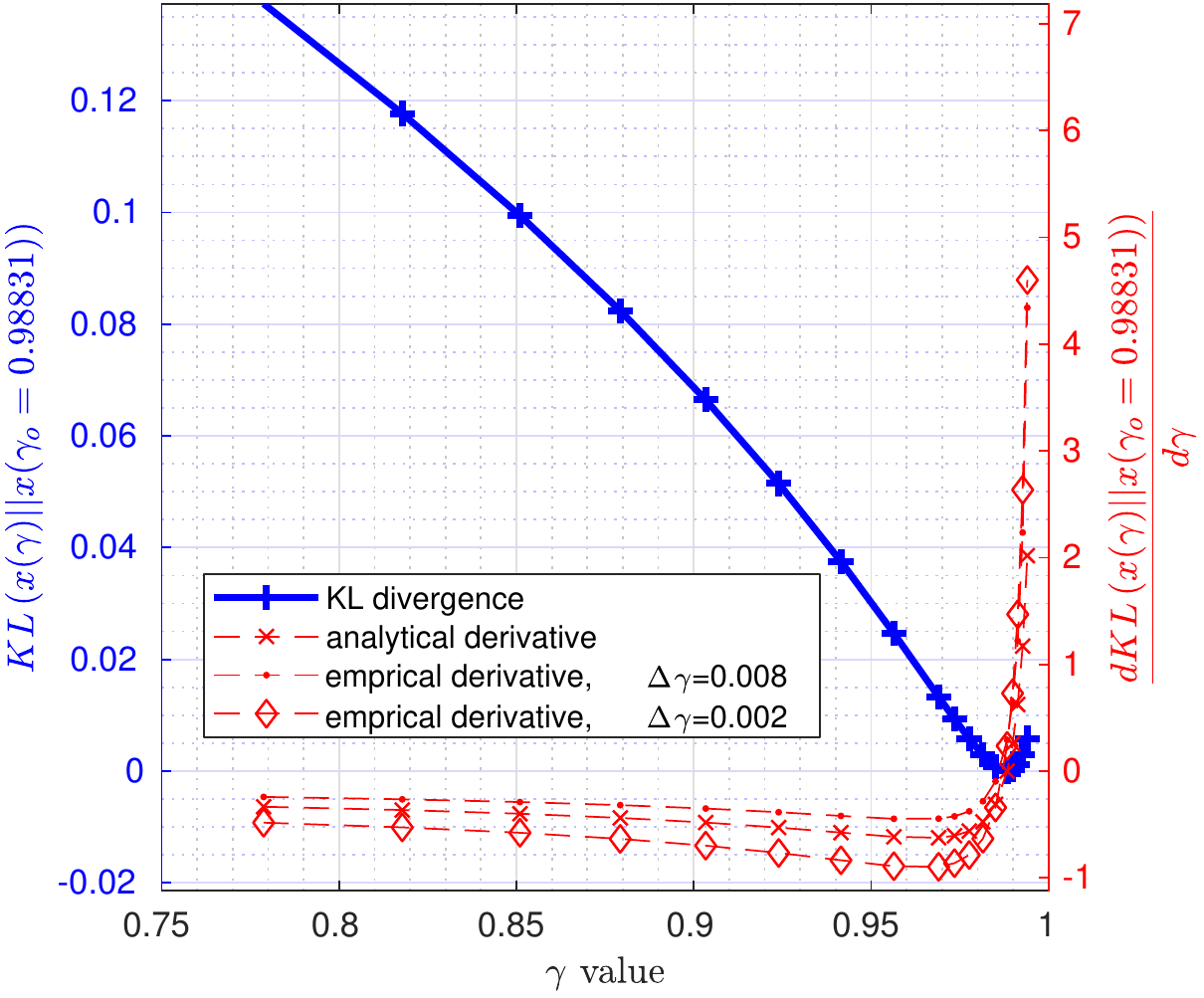}
  }
  \subfigure[{\footnotesize $\beta_o=19$}]{
    \includegraphics[width=0.14\textwidth]
          {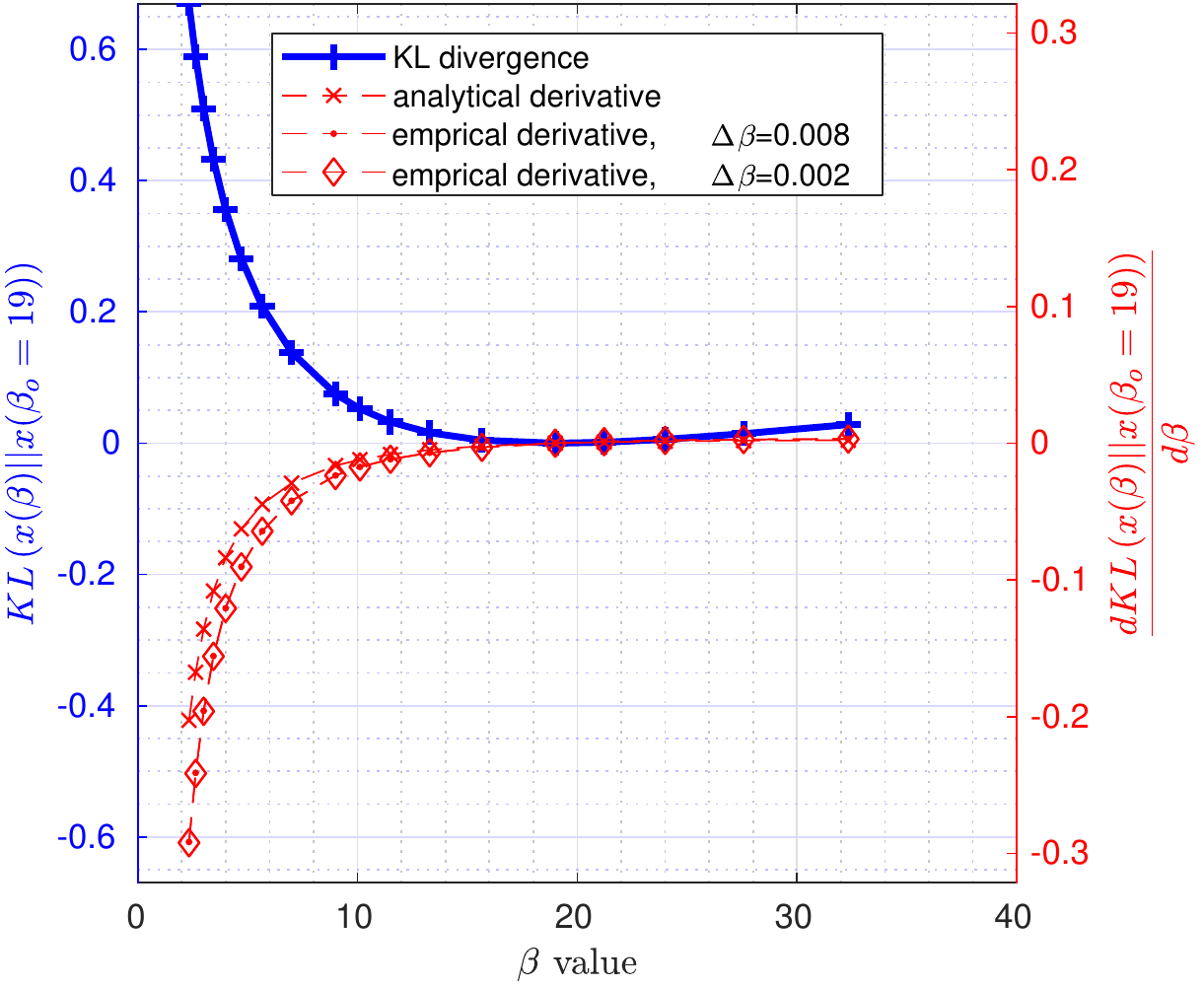}
  }
  
  \caption{\footnotesize Intra-model relative variation as defined in (\ref{eqn:kl}) in PgeRank
    distribution, on the Google graph, with respect to two reference
    distribution at $\alpha_{o} \in \{0.85,0.95\}$ with Brin-Page model ({\bf left
    column}), $\gamma_{o} \in \{0.94146,0.98831\}$ with log-$\gamma$ model ({\bf middle column}), 
    and $\beta_{o} \in \{5.\dot{6},19\}$ with Chung's model ({\bf right column}). 
    {\bf Blue curves:} the KL score 
    $KL( x_f(\rho) || x_{f}(\rho_{o}) )$ with numerically 
    computed distribution vectors; 
    {\bf Red curves:} the derivative of the KL score 
    $ (d/d\rho) KL( x_f(\rho) || x_{f}(\rho_{o} ))$. 
    The red curves with $\cdot$ marker and $\diamond$ marker are obtained empirically 
    from numerical distribution vectors, with step size 
    $\Delta \rho = (0.002, 0.008) $ respectively. The red curves with $\times$ marker
    are obtained analytically by (\ref{eqn:kl-derivative}). 
    {\bf Remarks.} With Brin-Page model and log-$\gamma$ model, the distribution changes gently from
    the reference distribution in the neighborhood of the reference
    value $\alpha_{o} = 0.85$, by the KL curve and the KL derivative
    curve.  In sharp contrast, the distribution with Chung's model
    deviates rapidly from the reference distribution.  }
  \label{fig:google-param-KL-0.85}
\end{figure}



\begin{figure}[!htb]
  \centering
  \subfigure[{\footnotesize $\alpha_o=0.85$}]{
    \includegraphics[width=0.14\textwidth]
          {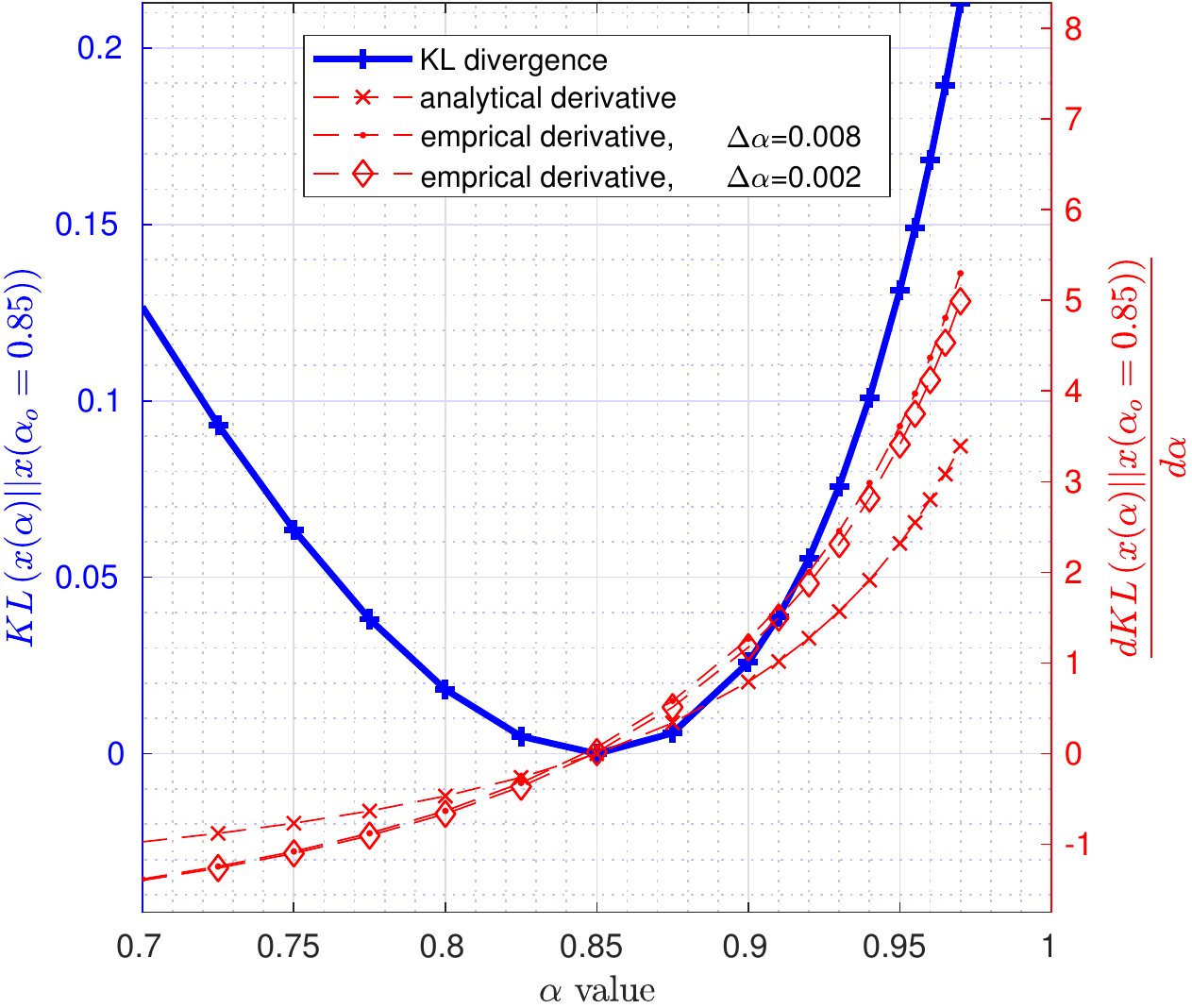}
  }
  \subfigure[{\footnotesize $\gamma_o=0.94146$}]{
    \includegraphics[width=0.14\textwidth]
          {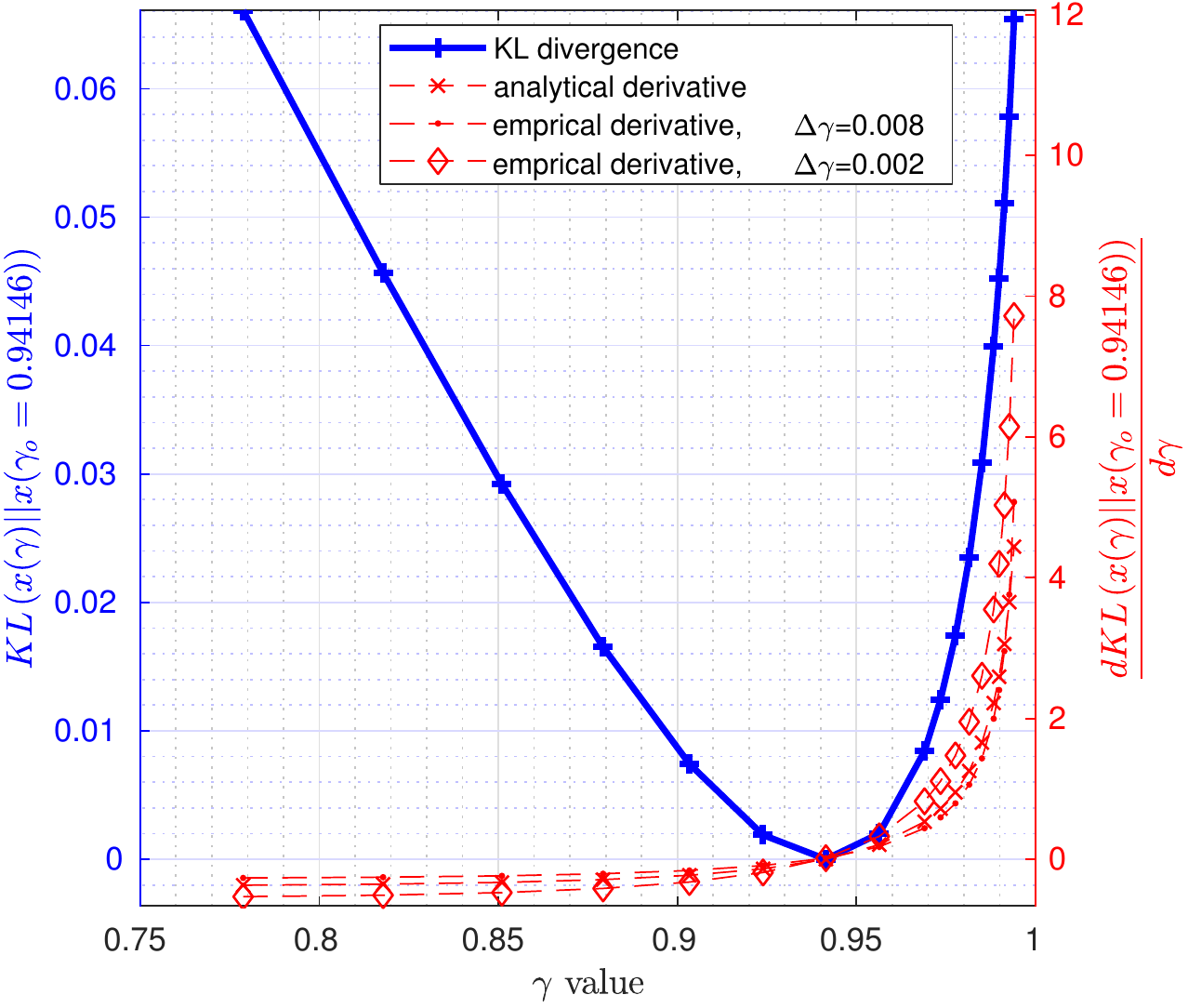}
  }
  \subfigure[{\footnotesize $\beta_o=5.\dot{6}$}]{
    \includegraphics[width=0.14\textwidth]
          {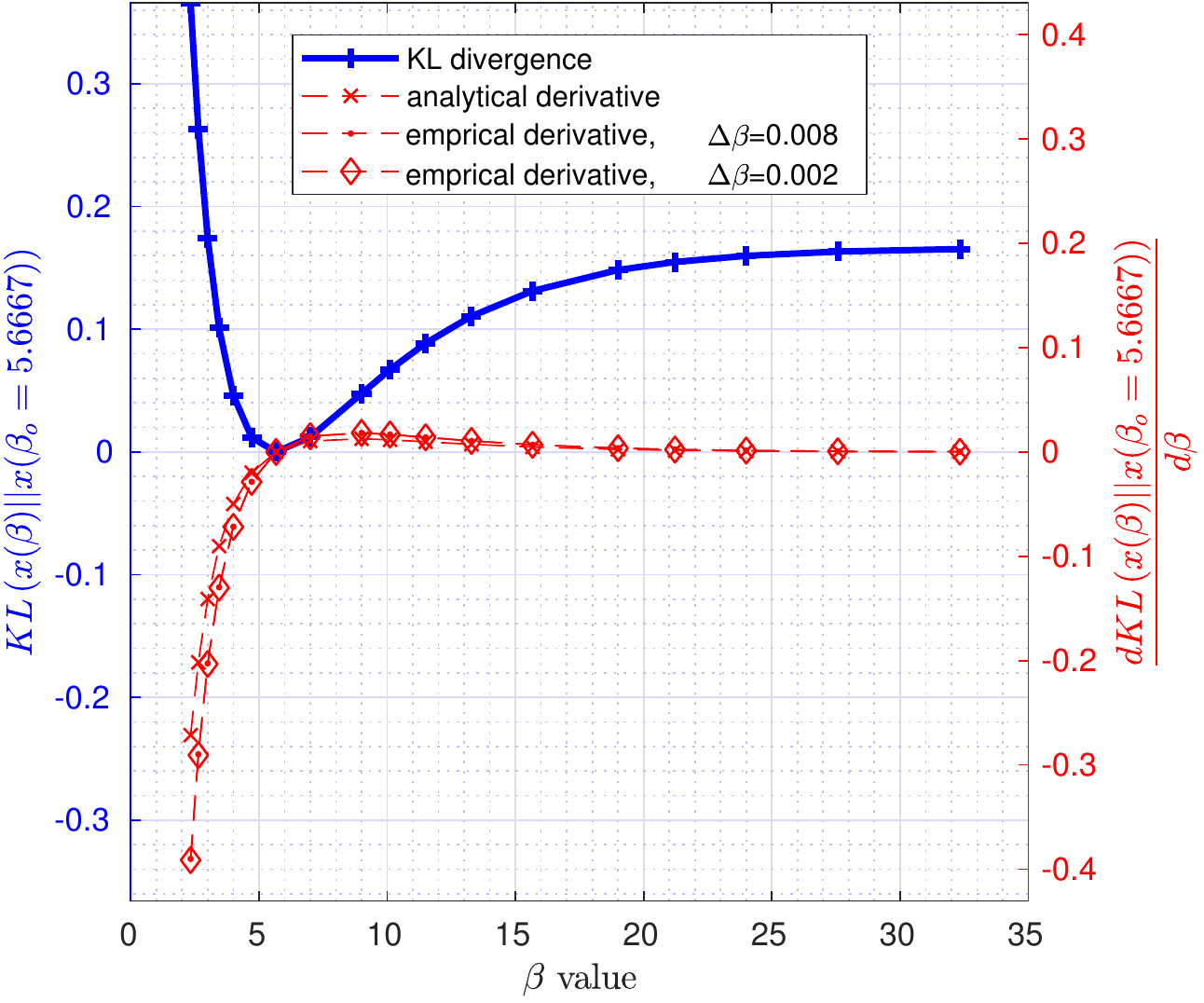}
  }
  \subfigure[{\footnotesize $\alpha_o=0.95$}]{
    \includegraphics[width=0.14\textwidth]
          {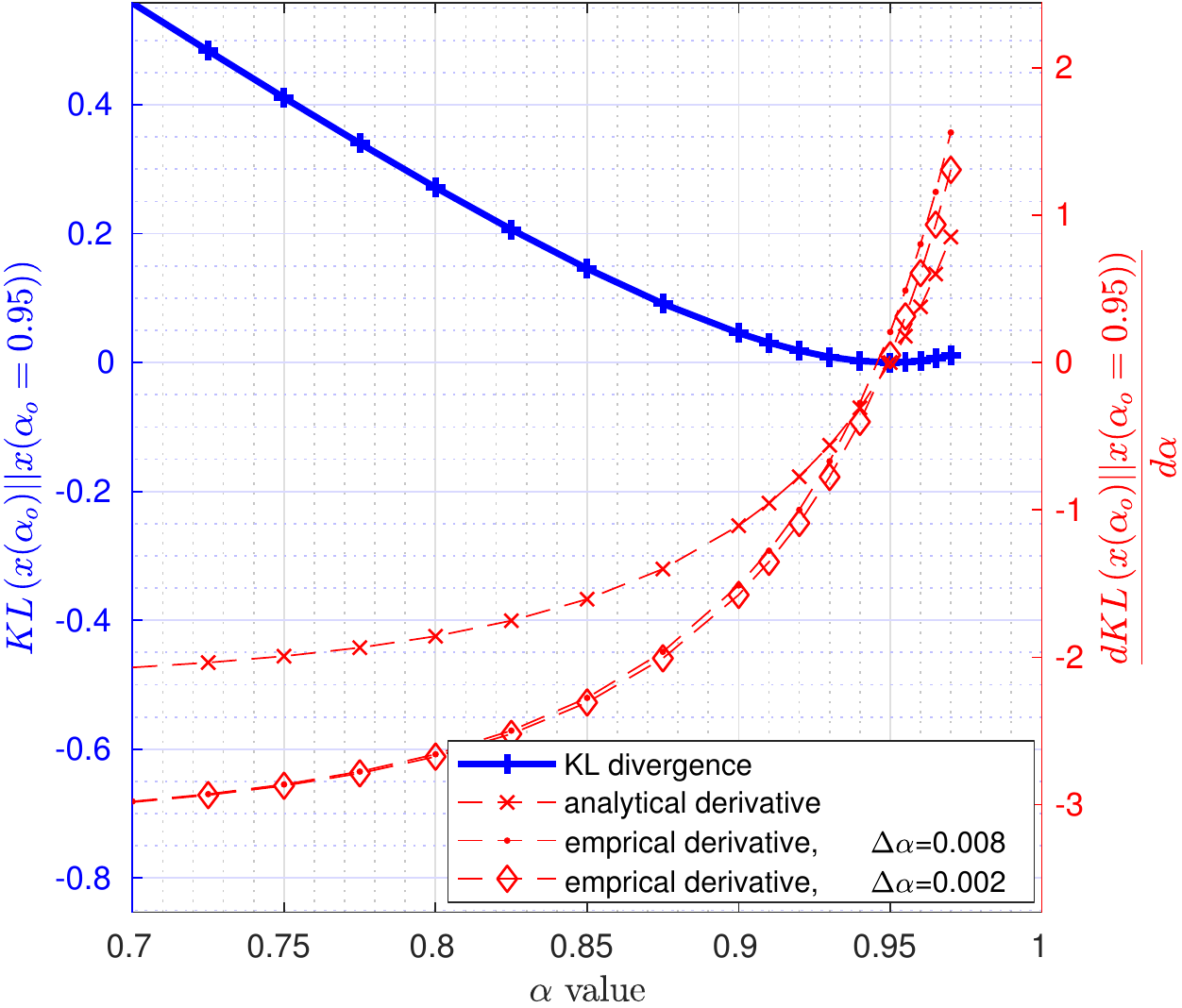}
  }
  \subfigure[{\footnotesize $\gamma_o=0.98831$}]{
    \includegraphics[width=0.14\textwidth]
          {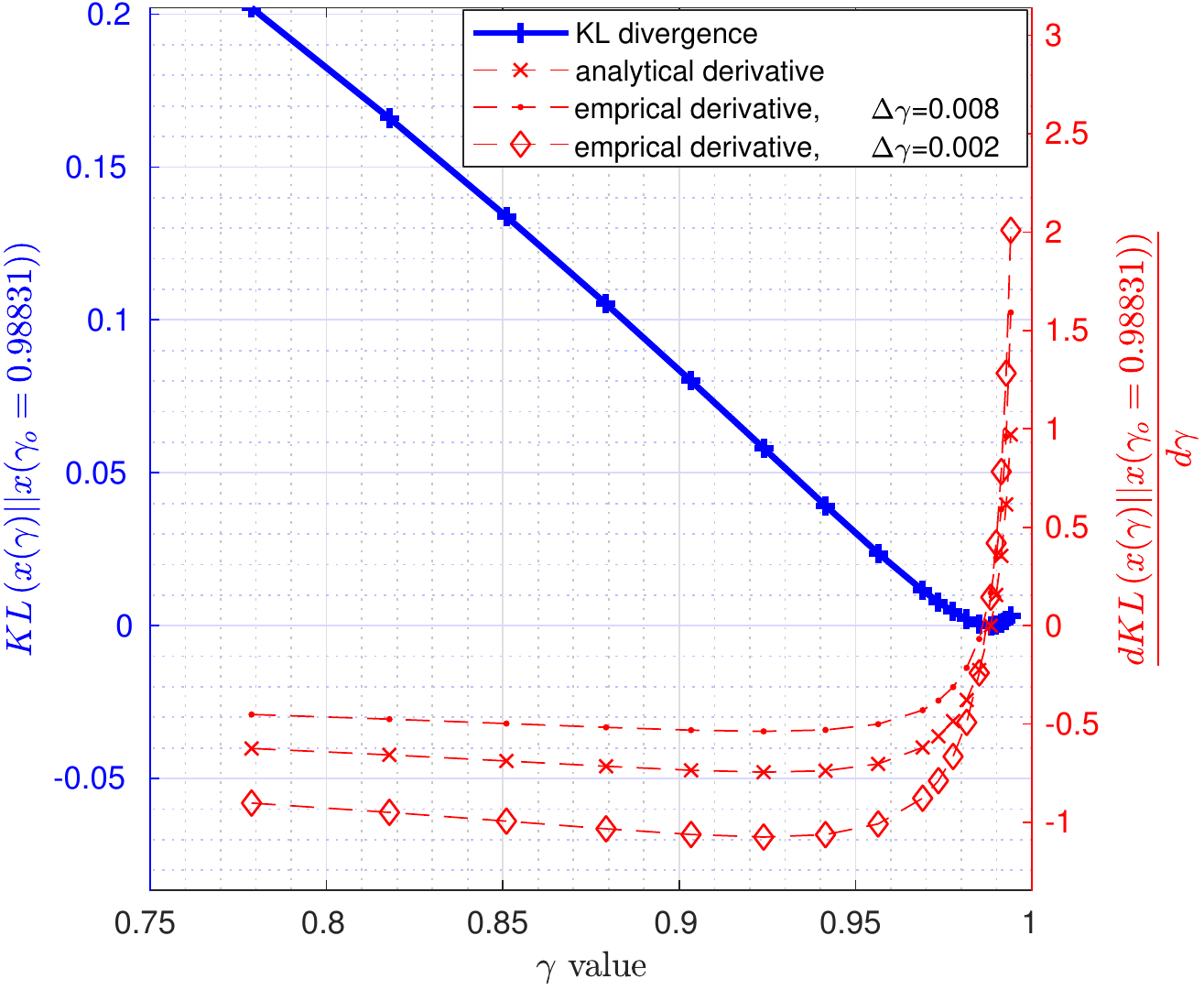}
  }
  \subfigure[{\footnotesize $\beta_o=19$}]{
    \includegraphics[width=0.14\textwidth]
          {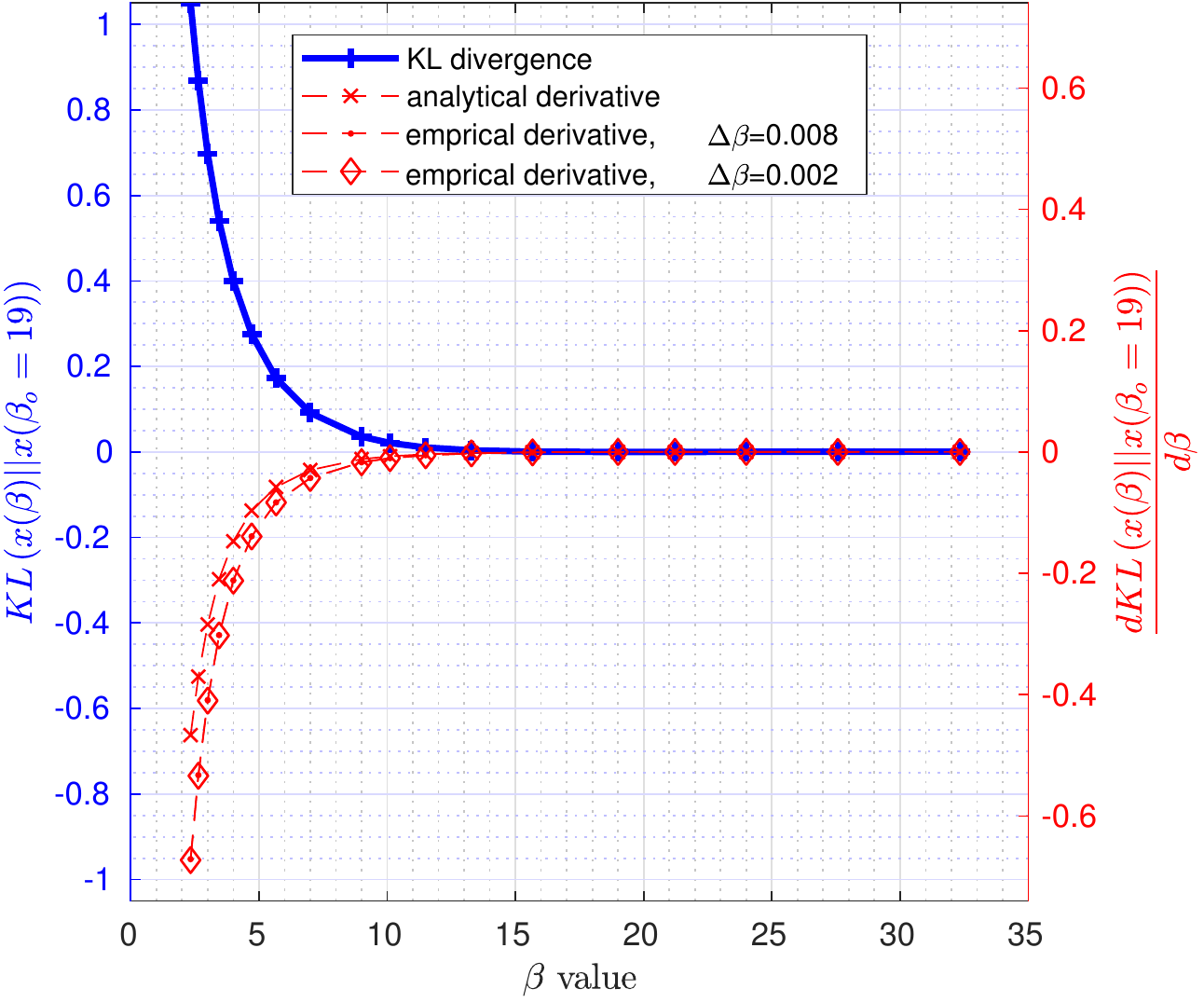}
  }
  \caption{\footnotesize Intra-model relative variation in PageRank
    distribution by (\ref{eqn:kl}), on the Twitter graph. The rest is in the same setting as in Figure \ref{fig:google-param-KL-0.85}.}
\label{fig:twitter-param-KL-0.85}
\end{figure}

We show the relative variation in PageRank vector with respect to a
reference vector by (\ref{eqn:kl}).  For Brin-Page model, we consider
two particular reference vectors: one is associated with $\alpha=0.85$
as chosen originally by Brin and Page, the other is at $\alpha = 0.95$,
much closer to the extreme case $\alpha=1$, in which the walks follow
the links only.  For Chung's model and log-$\gamma$ model, we use the corresponding parameter
values by (\ref{eqn:param-corres}).
We show the differences between the three models on the Google graph in
Figure~\ref{fig:google-param-KL-0.85} at the corresponding reference values, 
respectively, and on the twitter graph in
Figure~\ref{fig:twitter-param-KL-0.85}.

%

 
%

\subsection{Batch calculation: efficiency and accuracy} 
\label{subsec:batch-calculation} 
%
%

\begin{figure}[!htb]
  \centering
  \subfigure[Google]{
    \includegraphics[width=0.13\textwidth]{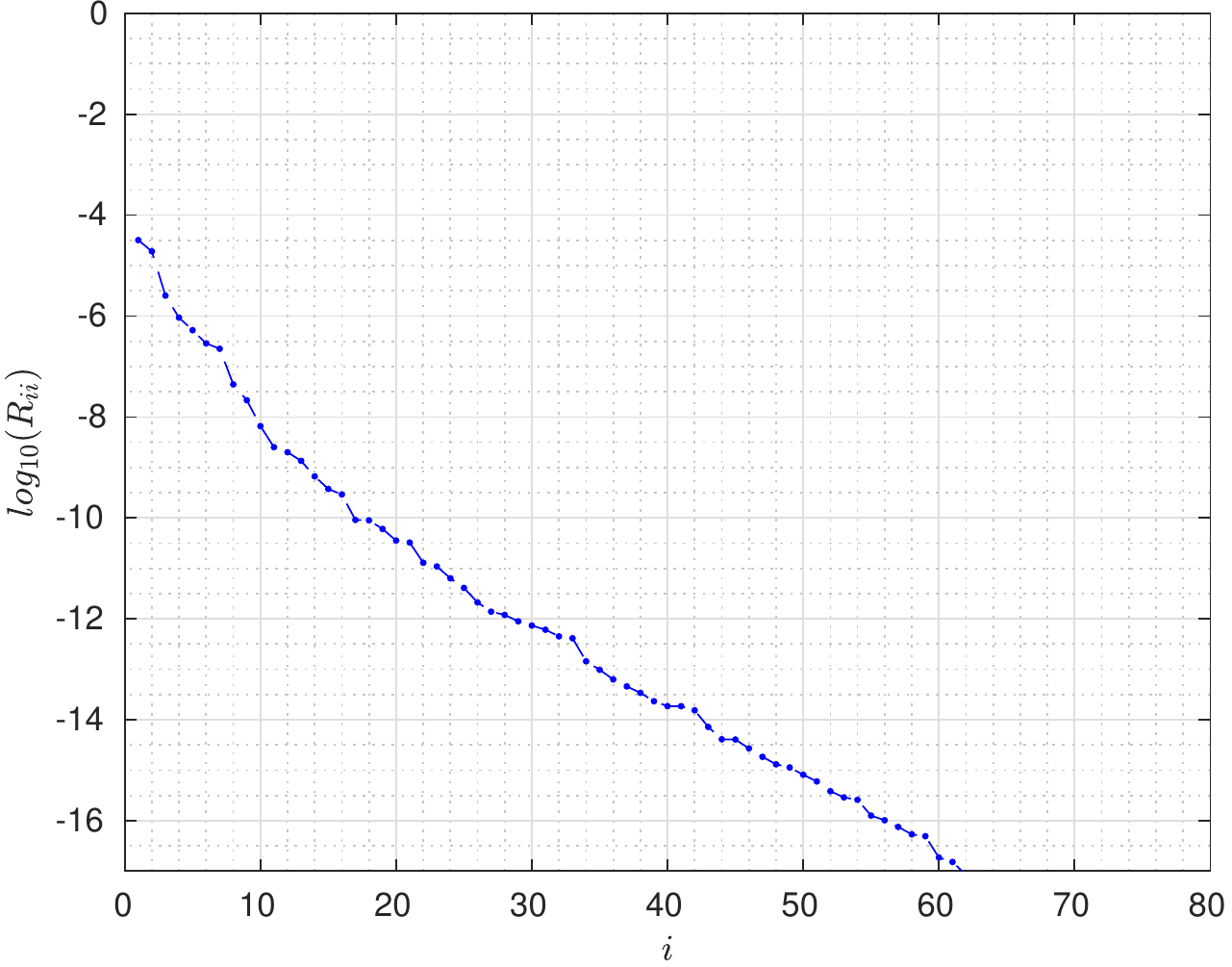}
  }
  \subfigure[Twitter(www)]{
    \includegraphics[width=0.13\textwidth]{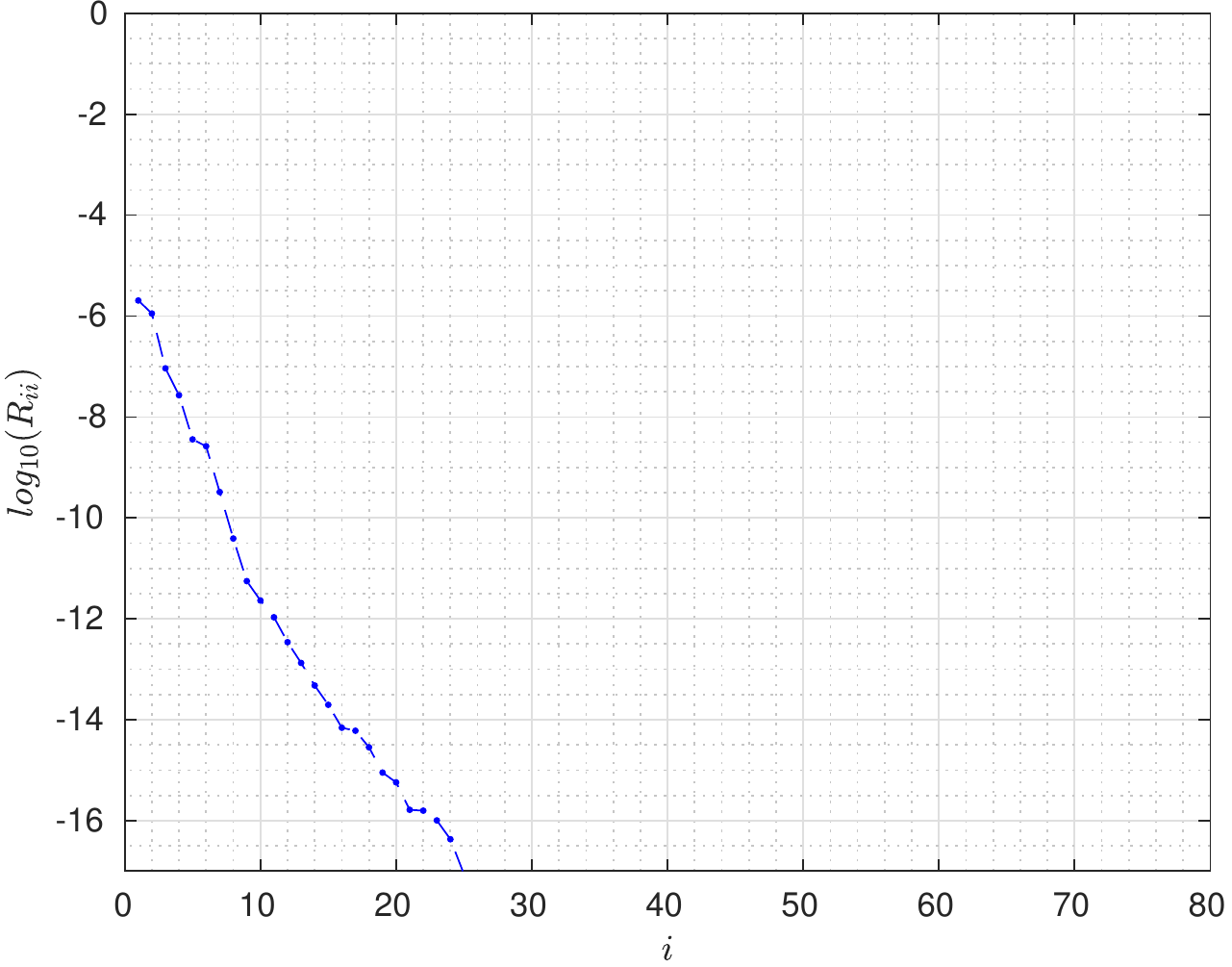}
  }
  \subfigure[Wikipedia]{
    \includegraphics[width=0.13\textwidth]{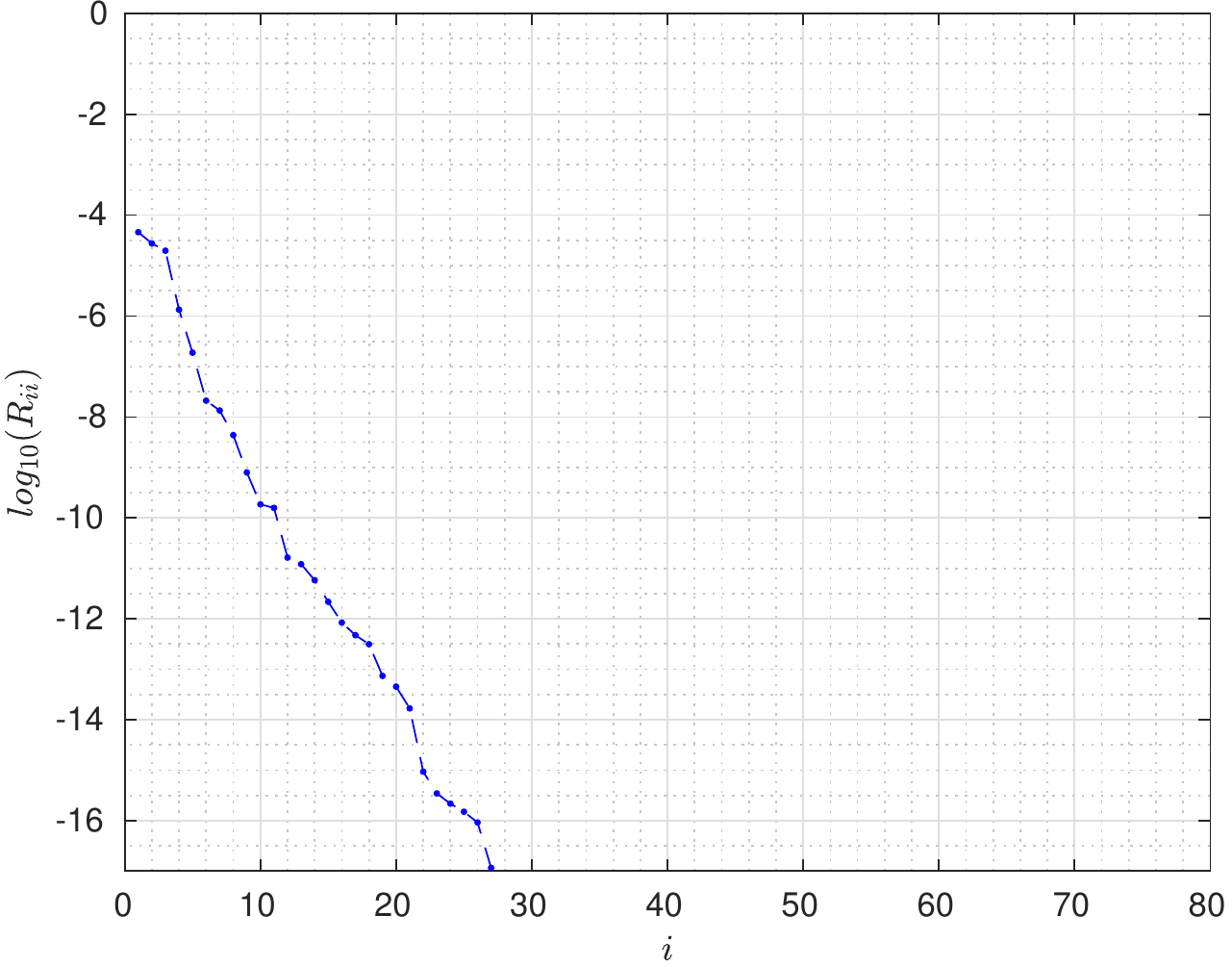}
  }
  \subfigure[DBpedia]{
    \includegraphics[width=0.13\textwidth]{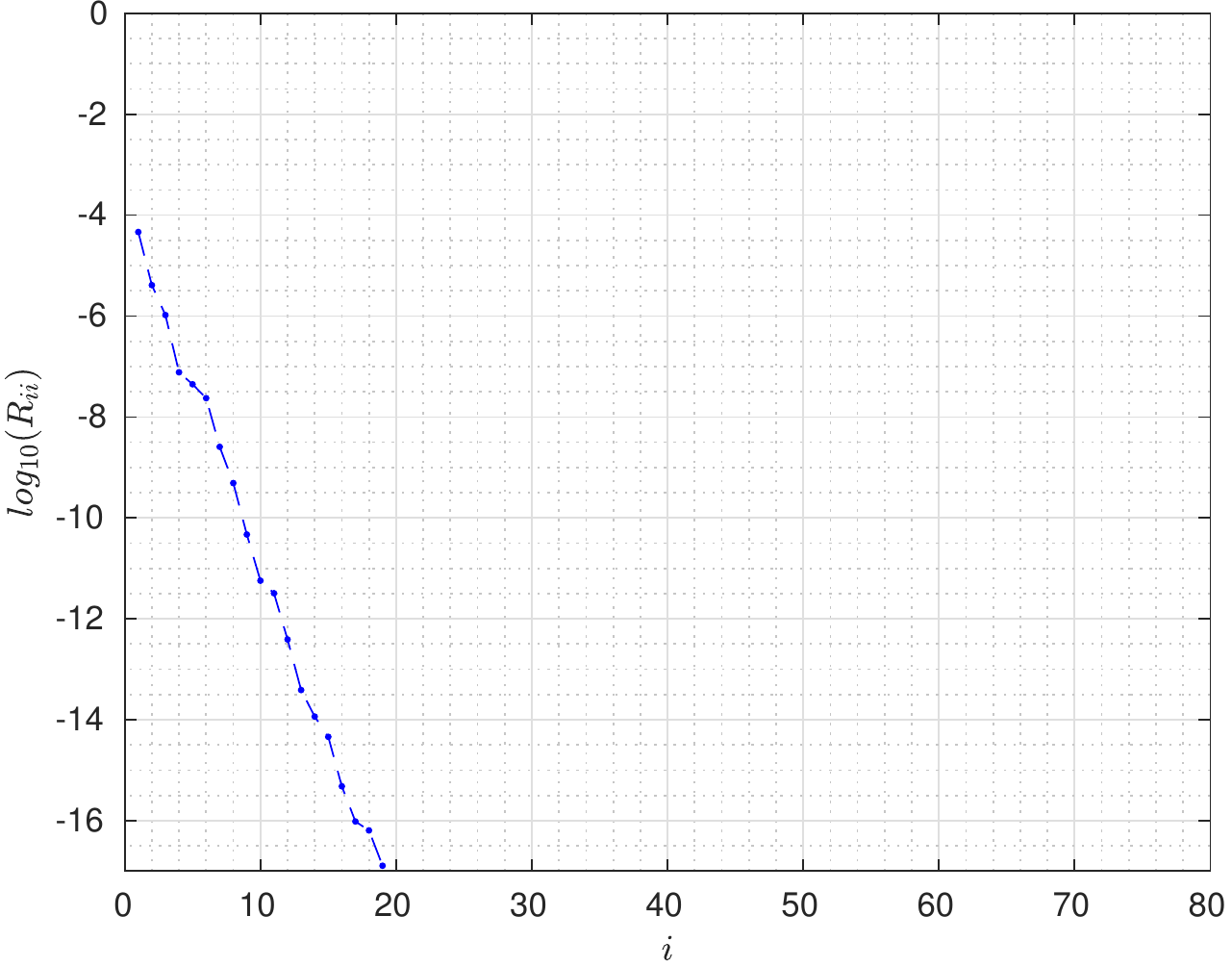}
  }
  \subfigure[Twitter(mpi)]{
    \includegraphics[width=0.13\textwidth]{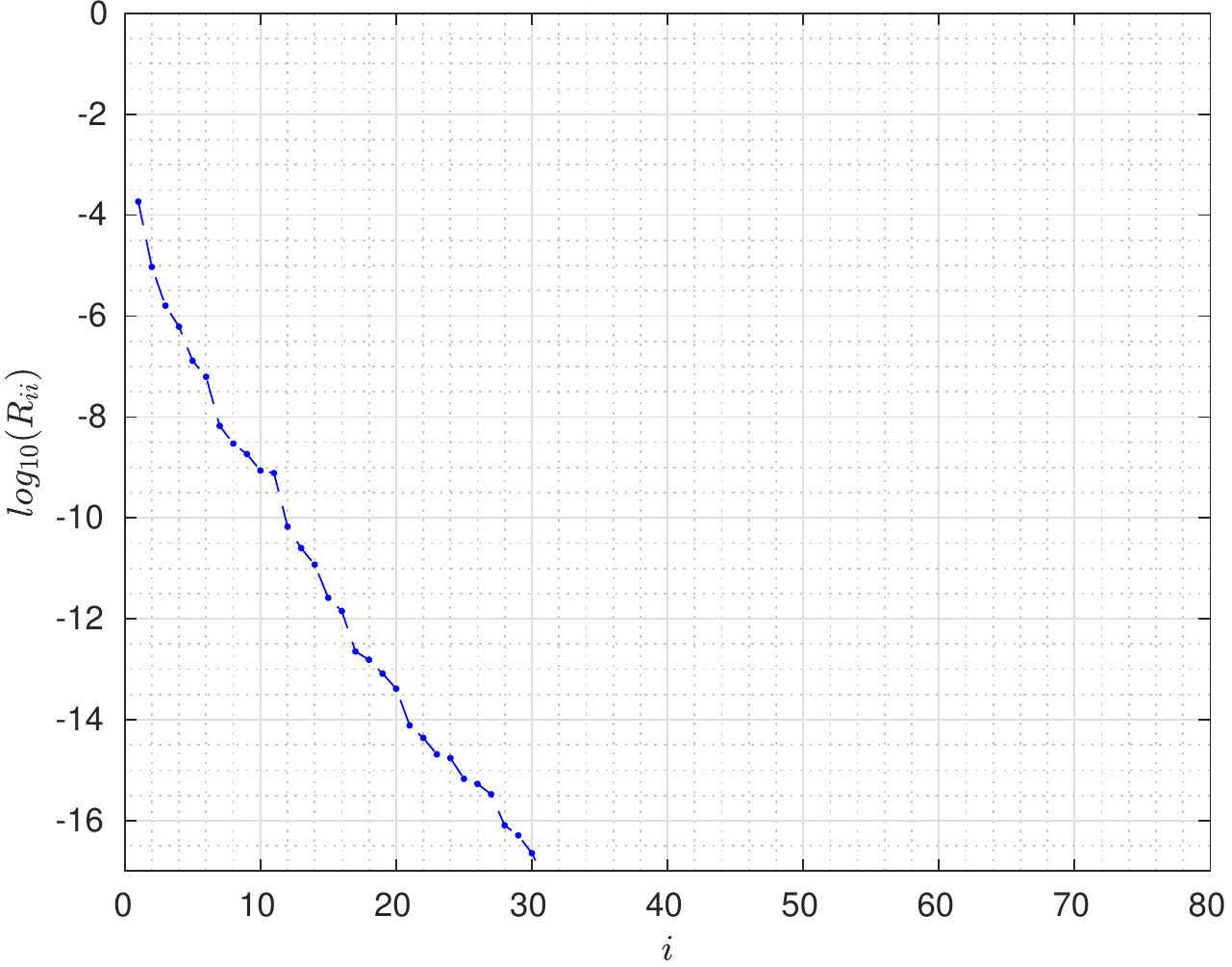}
  }
  \subfigure[Friendster]{
    \includegraphics[width=0.13\textwidth]{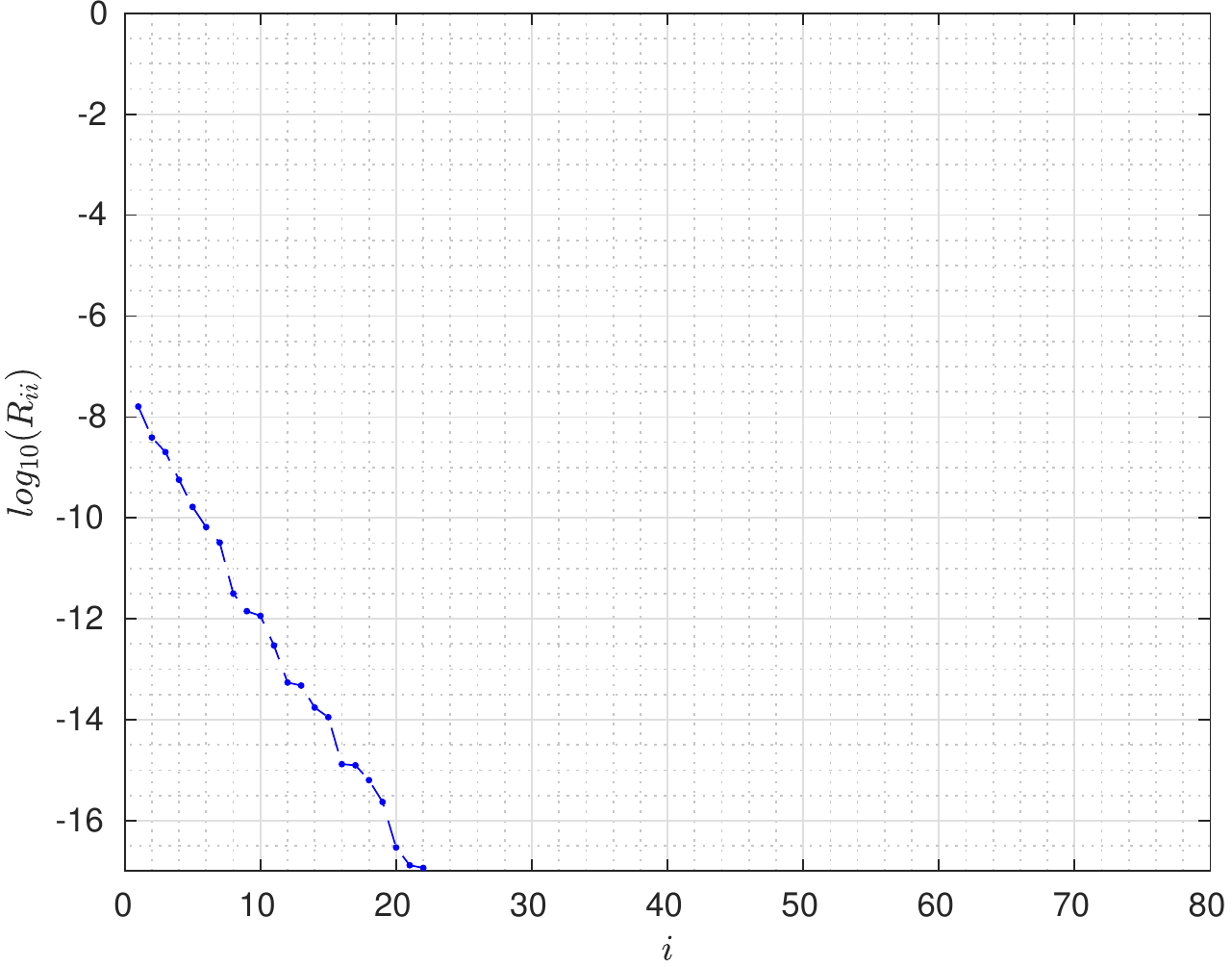}
  }
  \caption{The diagonal elements of each upper-triangular matrix $R$ obtained by a rank-revealing QR factorization for the 6 datasets in Table \ref{tab:dataset}. Google link graph has the highest numerical dimension 62 among the 6 datasets, and the DBpedia link graph has the lowest numerical dimension 19.}
  \label{fig:qr}
\end{figure}

We show first that the Krylov space dimension is numerically low for each of the 6 real world link graphs.
Figure \ref{fig:qr} gives the diagonal elements of each upper-triangular matrix $R$ obtained by a 
rank-revealing QR factorization. The elements below $10^{-17}$ are not shown. The numerical 
dimension ranges from $19$ with DBpedia link graph to $62$ with Google link graph. The low numerical 
dimension makes our algorithm in Section \ref{sec:batch-ranking} highly efficient. In addition, we 
exploited the sparsity of matrix $P$ in the Krylov vector calculation, see details in \cite{MS-Xichen-2018}.

%

The accuracy of our batch algorithm is evaluated in two ways. One is by 
$err = \|(x_{\text{Krylov}} - x_{\text{G-S}}) ./ x_{\text{G-S}}\|_{\infty}$, 
the maximum element-wise relative difference in
the PageRank vectors of Brin-Page model between the Gauss-Seidel method and 
our Krylov subspace method. In our experiments, the relative errors for all 6 datasets 
are below $10^{-10}$. In the other way, we show in 
Figure \ref{fig:google-param-KL-0.85} and Figure \ref{fig:twitter-param-KL-0.85} 
that the empirical rate of change agrees well with analytical prediction 
(\ref{eqn:kl-derivative}).

 
%


%


\section{Concluding remarks} 
%

Our model extension, connection, unified analysis and numerical
algorithm for quantitative estimation in batch are original, to our
knowledge. Our study leads to new observation and several findings.
\begin{inparaenum}[(a)]
\item In network propagation pattern in response to variation in the 
  damping mechanism, the inter-model difference among the 3 models
  is much more significant than the inter-dataset difference among
  the 6 datasets. This suggests the utility of
  model variety for differentiating network activities or propagation
  patterns.
\item The model solutions reside in the same customized, spectrally
  invariant subspace. On each of the $6$ real-world graphs, the space
  dimension is low, which is a small-world phenomenon.
\item The shared computation is not limited to one personalized
  distribution. The Krylov space associated with a particular vector $v$
  contains certainly many other distribution vectors. In fact, every
  Krylov vector is a distribution vector. This finding may lead to a much 
  more efficient way to represent and compute PageRank distributions 
  across multiple personalized vectors.
\item The low spectral dimension, estimated once for a particular graph
  $P$ and a personalized/customized vector $v$, may serve as a
  reasonable upper bound on the number of iterations by any competitive
  algorithm, with one matrix-vector product per iteration, for Brin-Page
  model, at any $\alpha$ value in $(0, 1)$, or any
  other model in the family (\ref{eqn:uni-random-walk}). The power method and the
  Gauss-Seidel iteration take more iterations to reach the same error
  level on the larger real-world graphs among the studied, and take many
  more iterations when $\alpha$ gets closer to $1$.
\end{inparaenum} 
In brief conclusion, estimating PageRank distribution under various
damping conditions is valuable and easily affordable.


%


\nocite{*}
\ifdefined\hpec

\else

\fi

\end{document}